\documentclass[twocolumn]{emulateapj}
\usepackage{ amsmath, color, natbib,graphicx,tablefootnote,subfigure}
\usepackage{ stmaryrd }

\newcommand\ii{{\sc ii}}
\newcommand\iii{{\sc iii}}

\newcommand\oiii{[O\,{\sc iii]}}
\newcommand\oii{[O\,{\sc ii]}}
\newcommand\siii{[S\,{\sc iii]}}
\newcommand\nii{[N\,{\sc ii]}}
\newcommand\sii{[S\,{\sc ii]}}

\shorttitle{CHAOS III}
\shortauthors{Croxall et al.}
\begin{document}  
\title{CHAOS III: Gas-Phase Abundances in NGC\,5457}
\author{
            Kevin V.\ Croxall$^{1}$, 
            Richard W.\ Pogge$^{1,2}$,  
            Danielle A.\ Berg$^{3}$, 
            Evan D.\ Skillman$^{4}$, 
            John Moustakas$^{5}$
}
\affil{
           $^1$Department of Astronomy, The Ohio State University, 140 W 18th Ave., Columbus, OH, 43210 \\
           $^2$Center for Cosmology \& AstroParticle Physics, The Ohio State University, 191 West Woodruff Avenue, Columbus, OH 43210 \\
           $^3$Center for Gravitation, Cosmology and Astrophysics, Department of Physics, University of Wisconsin Milwaukee, 1900 East Kenwood Boulevard, Milwaukee, WI 53211, USA \\
           $^4$Minnesota Institute for Astrophysics, University of Minnesota, 116 Church St. SE, Minneapolis, MN 55455 \\
           $^5$Department of Physics \& Astronomy, Siena College, 515 Loudon Road, Loudonville, NY 12211\\
}

\begin{abstract}
The CHemical Abundances of Spirals (CHAOS) project leverages the combined power of the Large Binocular Telescope (LBT) with the broad spectral range and sensitivity of the Multi Object Double Spectrograph (MODS) to measure ``direct'' abundances (based on observations of the temperature-sensitive auroral lines) in large samples of \ion{H}{2} regions in spiral galaxies.  We present LBT MODS observations of 109 H\ii\ regions in NGC\,5457 (M\,101), of which 74 have robust measurements of key auroral lines (50 [\ion{O}{3}] $\lambda4363$, 47 [\ion{N}{2}] $\lambda5755$, 59 [\ion{S}{3}] $\lambda6312$,  67 [\ion{O}{2}] $\lambda\lambda7320,7330$ and  70 [\ion{S}{2}] $\lambda4076$ at a strength of $\ge$ 3 $\sigma$), a factor of $\sim$3 larger than all previous published detections of auroral lines in the H\ii\ regions of NGC\,5457. Comparing the temperatures derived from the different ionic species we find: (1) strong correlations of T[N\,\ii] with T[S\,\iii] and T[O\,\iii], consistent with little or no intrinsic scatter; (2) a correlation of T[S\,\iii] with T[O\,\iii], but with significant intrinsic dispersion; (3) overall agreement between T[N\,\ii], T[S\,\ii], and T[O\,\ii], as expected, but with significant outliers; (4) the correlations of T[N\,\ii] with T[S\,\iii] and T[O\,\iii] match the predictions of photoionization modeling while the correlation of T[S\,\iii] with T[O\,\iii] is offset from the prediction of photoionization modeling. Based on these observations, which include significantly more observations of lower excitation H\ii\ regions, missing in many analyses, we inspect the commonly used ionization correction factors (ICFs) for unobserved ionic species and propose new empirical ICFs for S and Ar. We have discovered an unexpected population of \ion{H}{2} regions with a significant offset to low values in Ne/O, which defies explanation. We derive radial gradients in O/H and N/O which agree with previous studies. Our large observational database allows us to examine the dispersion in abundances, and we find intrinsic dispersions of $0.074\pm0.009$ in O/H and $0.095\pm0.009$ in N/O  (at a given radius).  We stress that this measurement of the intrinsic dispersion comes exclusively from direct measurements of H\ii\ regions in NGC\,5457.  
\end{abstract} 
 
 
%%%%%%%%%%%%%%%%%%%%%%%%%%%%%%%%%%%%%%%%%%%%%%%%%%%%%%%%%%%%%%%%
 \section{Introduction}
 \subsection{CHAOS}
 As the dominant sites of current star formation, nearby spiral galaxies provide the best laboratories for understanding the star formation process.  Presently we are secure in our knowledge of several properties of present day spiral galaxies.  Over the past decades, the morphological sequence identified by Hubble and contemporaries has become understood as a sequence in both mass and gas content \citep[e.g.,][]{blanton2009}. The more actively star forming galaxies are generally of lower mass and higher gas content, indicating that the efficiency of star formation is related to the mass surface density of galaxies. We also know that the vast majority of spiral galaxies show a correlation between their luminosities and the abundance of heavy elements in their interstellar media \citep[e.g.,][]{shields1990}. The dispersion in this correlation is markedly smaller when one plots heavy element abundance versus stellar mass, indicating that it is the mass of the galaxy that drives this correlation \citep[e.g.,][]{tremonti2004}.  Additionally, the chemical abundances of spiral galaxies are known to show radial gradients \citep[e.g.,][]{pageledmunds1981}. While this first-order description of present day spiral galaxies is secure, there are important aspects, in particular regarding their chemical evolution, that are, simply put, unknown. These missing details are preventing us from understanding key processes in the evolution of galaxies. 

While numerous spectra of star forming galaxies have been obtained through large surveys such as SDSS \citep{tremonti2004}, PINGS \citep{pings}, CALIFA \citep{marino2013}, MaNGA \citep{bundy2015,law2015}, and SAMI \citep{bryant2015} few of these observations enable direct determinations of absolute gas-phase abundances, as they do not detect the faint auroral lines that reveal the electron temperatures of the H\ii\ regions.  Even determining the relative abundances can be challenging given possible biases, both in the observations and the methodology of determining gas-phase metallicity using only the brightest lines \citep{kewley2008}.  Furthermore, the coarse spatial resolution that results from observing distant galaxies means that non-homogenous clouds of gas will co-inhabit each spectrum \citep{moustakas2006}.  

Assembling very high quality spectra of H\ii\ regions in spiral galaxies, which allow accurate determinations of absolute and relative abundances, reveals the complex nature of understanding chemical evolution across a broad range of parameter space. Observations of HII regions in spiral galaxies indicate that multiple temperature diagnostics are useful when attempting to understand the ionization structure of nebulae \citep{berg2015}.  Failure to measure multiple ionic temperatures, in cases with single discrepant temperatures or systematic biases, can lead to an increased apparent dispersion in the chemical abundances at a given radius, and increased dispersions can result in an artificial flattening of gradients.

\begin{figure*}[tp] 
\epsscale{0.55}
   \centering
   \plotone{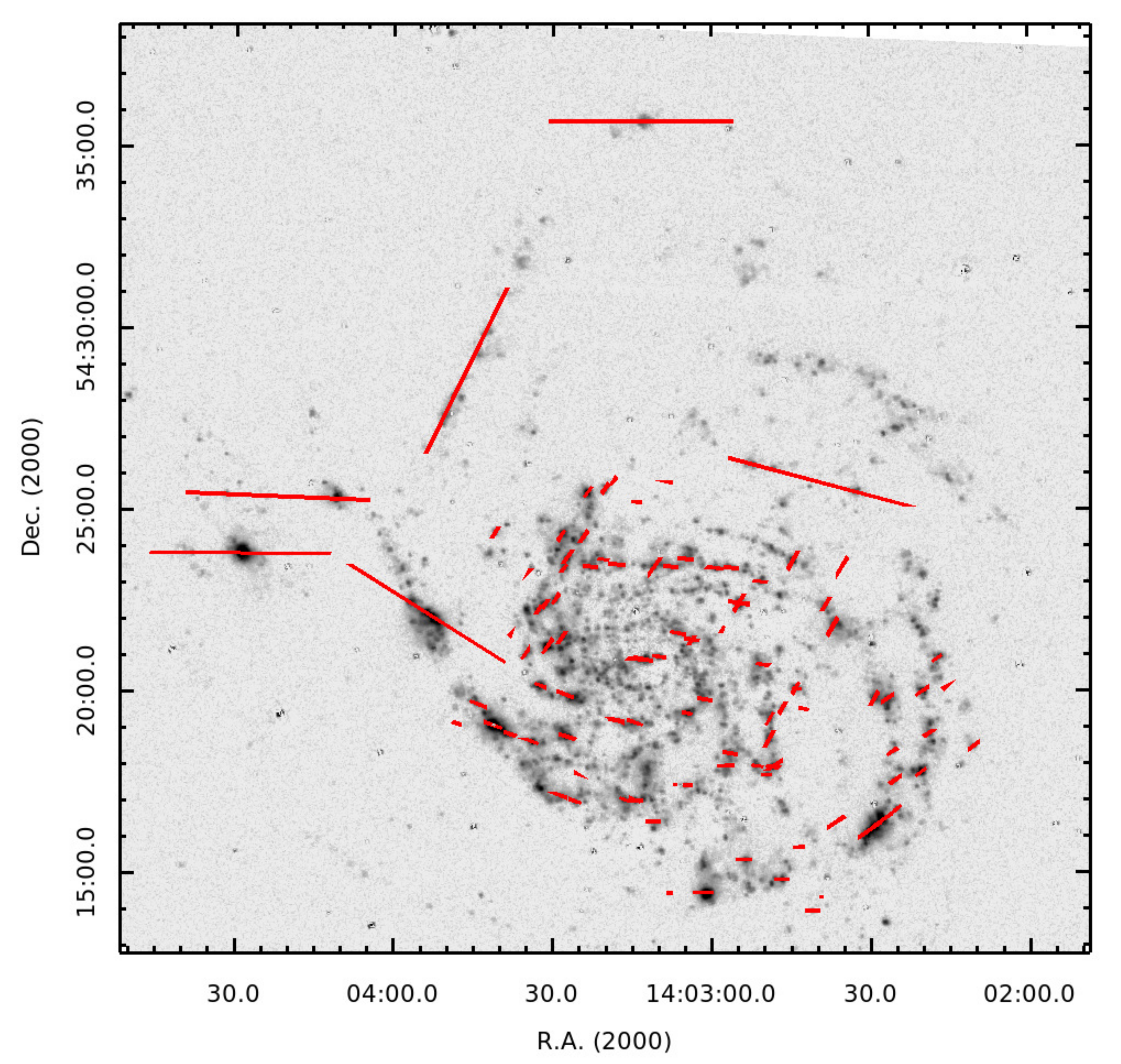}
   \plotone{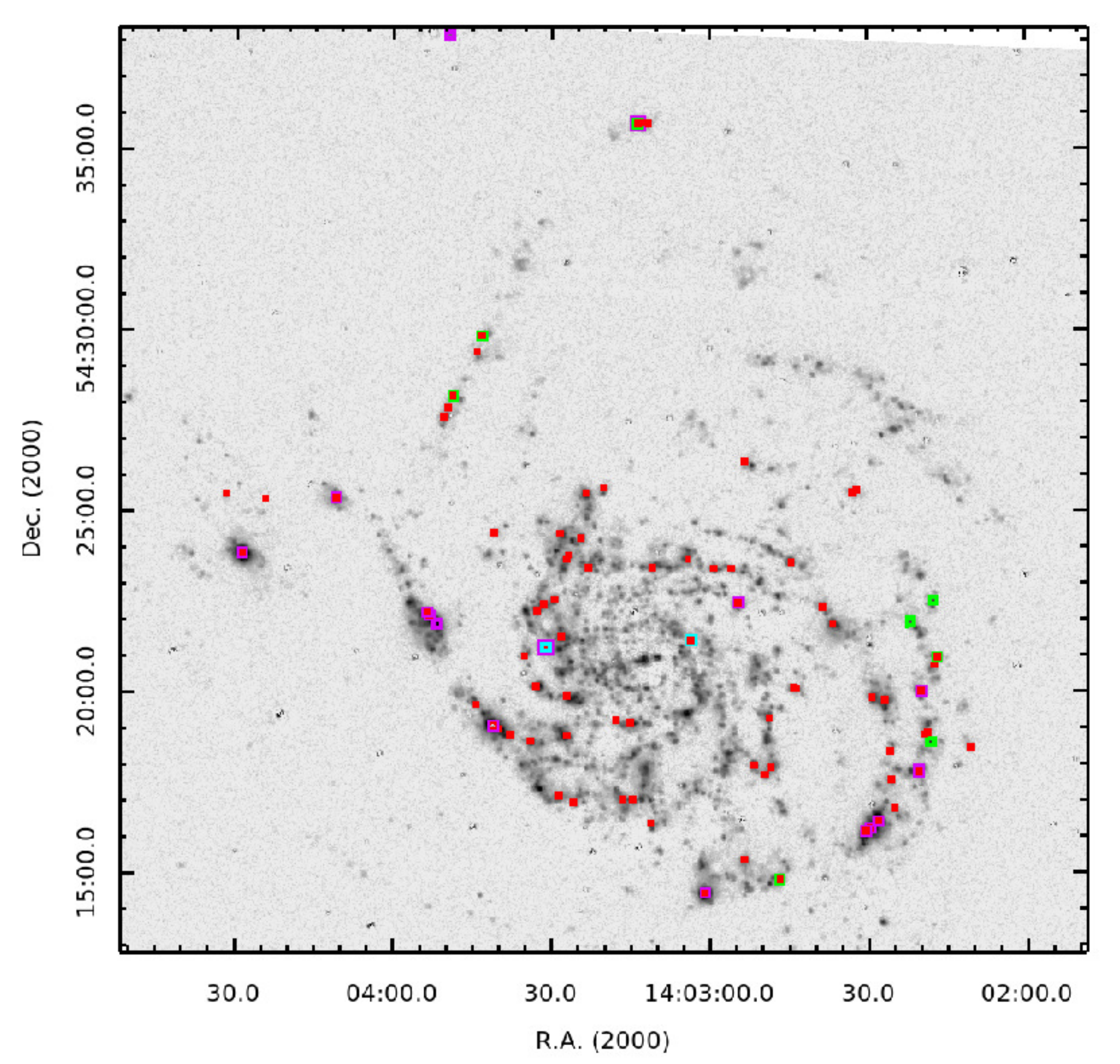}
   \caption{Map of targeted H\ii\ regions in NGC\,5457 overlaid on an H$\alpha$ image.  Offsets in arc seconds, relative to the galaxy center,  are given in Table \ref{t:locations}.  Left: Targeted slit locations of all CHAOS observations in NGC\,5457.  Right: Locations of H\ii\ regions where auroral lines were detected.  Magenta, Cyan, green and red denote observations reported in \citet{kennicutt2003}, \citet{bresolin2007}, \citet{li2013}, and this work respectively.\\}   
   \label{fig:slits}
\end{figure*}   

We recognize that having a direct abundance is not the final word on abundances and that systemic effects can exist for direct abundances also \cite[e.g.,][]{peimbert1967,stasinska2005}.  However, we believe that obtaining a very large sample of direct abundance is our best chance to assess the possible systematics in the direct abundances.

 \subsection{NGC5457 / M101}
The nearly face-on spiral galaxy NGC\,5457 (also known as M\,101) presents numerous luminous H\ii\ regions for observation.  In a seminal study, \citet{kennicutt2003} measured direct oxygen abundances in 20 H\ii\ regions spanning the optical disk of NGC\,5457, showing that this galaxy exhibits a steep metallicity gradient.  More recently, \citet{li2013} found indications of possible azimuthal abundance variations via observation of nine additional H\ii\ regions between 0.55$\leq$R/R$_{25}$$\leq0.6\,$kpc.  

While  \citet{kennicutt2003} found good agreement between various ionic temperatures and the photoionization models of \citet{garnett1992}, a re-analysis of the reported line fluxes from \citet{kennicutt2003}, using updated atomic data, find possible disagreements between these ionic temperatures \citep{berg2015}.  However, we note that the [O\,\iii]\,$\lambda$4363 and [N\,\ii]\,$\lambda$5755 lines were simultaneously detected in only 7 H\ii\ regions.  Including regions from \citet{li2013} does not clarify the situation, as their spectra do not extend redward of 5100\AA.  Furthermore, both of these studies primarily focus on metal poor H\ii\ regions where these discrepancies seem to diminish \citep{binette2012}.  Accordingly, we are able to use multiple emission line diagnostics to determine the physical conditions in the H\ii\ regions and trace disagreements between measured temperatures and photoionization models, with a set goal of determining absolute and relative chemical abundances with uncertainties $<$0.1 dex. 

Here we present observations taken with the Multi-Object Double Spectrographs \citep[MODS,][]{mods} on the Large Binocular Telescope (LBT).  With these observations we increase the number of direct oxygen abundance measurements in NGC\,5457 by a factor of $\sim$3.  This will enable a broader study of possible extra-radial variations in abundance.  

We have observed H\ii\ regions in NGC\,5457 as part of CHAOS to increase the number of auroral line detections, to investigate multiple means of determining the electron temperature in metal-rich galaxies, and to verify its absolute metallicity gradient.  Our observations and data reduction are described in \S2.  In \S3 we determine electron temperatures and direct gas-phase chemical abundances.  We present abundance gradients for O/H and N/O in \S4.  We discuss these abundance gradients and the azmuthal patterns of chemical evolution in \S5.   Finally, we summarize our conclusions in~\S6.
 
%%%%%%%%%%%%%%%%%%%%%%%%%%%%%%%%%%%%%%%%%%%%%%%%%%%%%%%%%%%%%%%%
\section {Observations}
\subsection{Optical Spectroscopy}
Optical spectra of NGC\,5457 were taken using MODS on the LBT during the spring semester of 2015.  All spectra were acquired with the MODS1 unit.  We obtained simultaneous blue and red spectra using the G400L (400 lines mm$^{-1}$, R$\approx$1850) and G670L (250 lines mm$^{-1}$, R$\approx$2300) gratings, respectively.  This setup provided broad spectral coverage extending from 3200\,--\,10,000 \AA.  In order to detect the intrinsically weak auroral lines, i.e., [O\,\iii] $\lambda$4363, [N\,\ii] $\lambda$5755, and [S\,\iii] $\lambda$6312, in numerous H\ii\ regions, 13 fields were targeted; seven using multi-object masks and six using the 1\farcs2 facility long-slit mask.    The individual field masks were cut such that $\sim$15--20 spectra, including both H\ii\ regions and sky-slits, were simultaneously obtained.   We obtained multiple exposures of 1200s for each field.  For multi-object masks we obtained six exposures, for a total integration time of 2-hours for each;  long-slit targets were typically luminous, massive H\ii\ regions and thus were only observed with between three and five exposures each, depending on the cloud cover at the time of observations.
\begin{figure*}[tbp] 
\epsscale{1.2}
   \centering
   \plotone{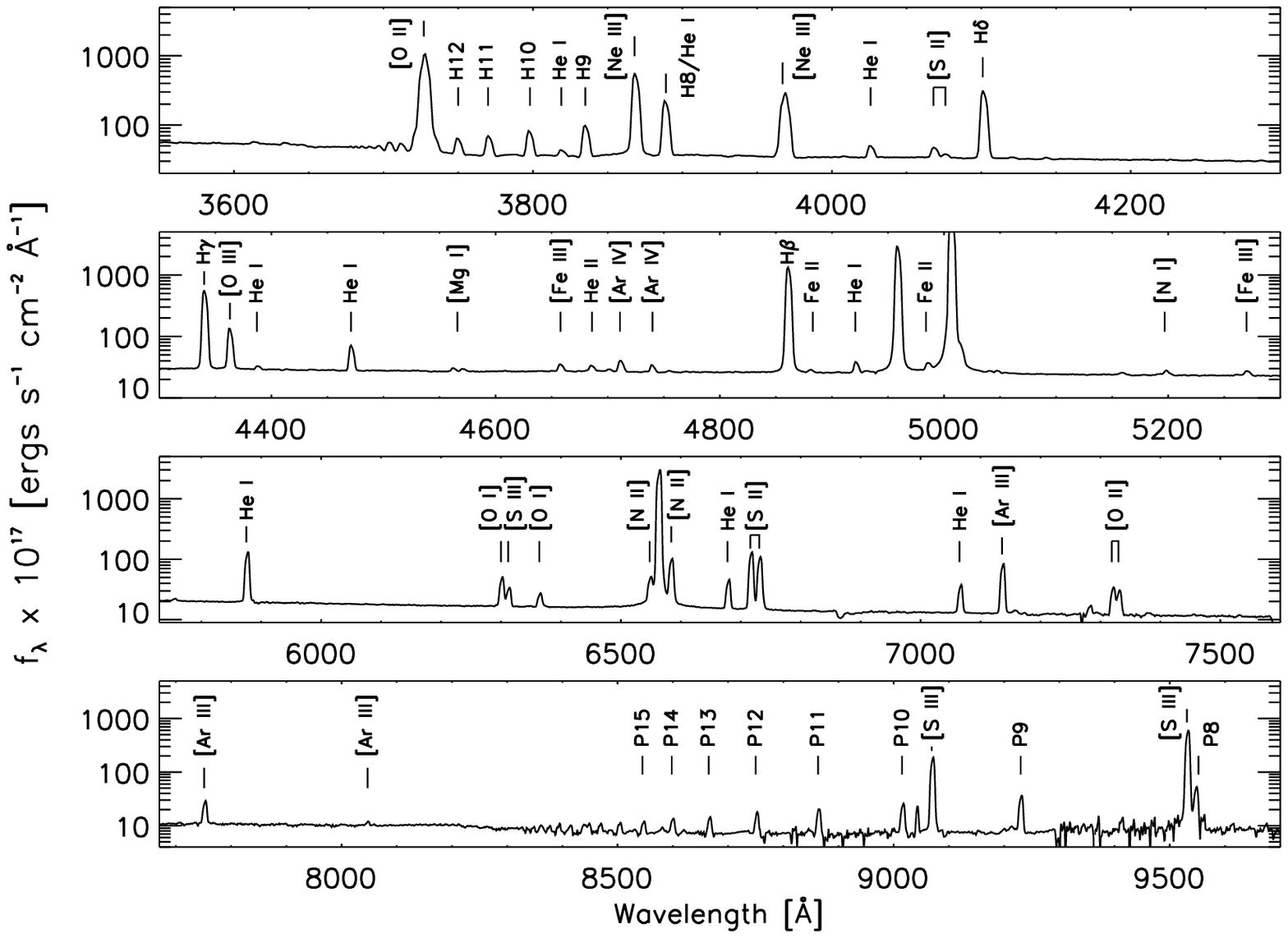}
   \caption{Example of a one dimensional spectrum taken from MODS1 observations of NGC\,5457, namely NGC5457+667.9+174.1. Notable major emission features are marked and labeled.  We have not corrected for major telluric absorption features.}
   \label{fig:spectra}
\end{figure*}  

Figure \ref{fig:slits} shows the locations of slits overlaid on an H$\alpha$ map of NGC\,5457.  Throughout this work, we label all locations as offsets, in right ascension and declination, from the center of NGC\,5457, as listed in Table \ref{t:n5457global}.  We obtained our observations when NGC\,5457 was at relatively low airmass, typically $\sim$1.2.  We also cut our slits close to the median parallactic angle of the observing window. The combination of low airmass and matching the parallactic angle minimizes flux lost due to differential atmospheric refraction between 3200 -- 10,000 \AA \citep[cf.][]{filippenko1982}.

Our targeted regions (see Table \ref{t:locations}), as well as alignment stars, were selected based on archival broad-band and H$\alpha$ imaging \citep{hoopes2001}.  We cut most slits to be $\sim$15\arcsec\ long with a 1\arcsec\ slit width.  Slits were placed on relatively bright H\ii\ regions across the entirety of the disk; this ensured that both radial and azimuthal  trends in the abundances could be investigated.  When extra space between slits was available, slits were extended to make the best use of the available field of view.  As the bright disk of NGC\,5457 complicates local sky subtraction, we also cut sky-slits in each mask that provided a basis for clean sky subtraction.

\begin{deluxetable}{lcccccccccc}  
\tabletypesize{\scriptsize}
\tablecaption{Adopted Global Properties of NGC\,5457}
\tablewidth{0pt}
\tablehead{ 
  \colhead{Property}	&
  \colhead{Adopted Value}	&
  \colhead{Reference}	
  }
\startdata
R.A. 	& 14$^h$03$^m$12.5$^s$ &  1 \\  
Dec  & +54$^\circ$20$^m$56$^s$ & 1 \\
Inclination & 18$^\circ$ & 2\\
Position Angle  & 39$^\circ$ & 2\\
Distance & 7.4 Mpc & 3 \\
R$_{25}$ & 864\arcsec & 4 
\enddata
\label{t:n5457global}
\tablecomments{Units of right ascension are hours, minutes, and seconds, and units of declination are degrees, arcminutes, and arcseconds. References are as follows: [1]  NED  [2] Walter et al. 2008 [3] Ferrarese et al. 2000 [4] Kennicutt et al. 2011}
\end{deluxetable}

For a detailed description of the data reduction procedures we refer the reader to \citet{berg2015}.  We only note here the primary points of our data processing.  Spectra were reduced and analyzed using the development version of the MODS reduction pipeline\footnote{http://www.astronomy.ohio-state.edu/MODS/Software/modsIDL/} which runs within the XIDL\footnote{http://www.ucolick.org/$\sim$xavier/IDL/} reduction package.  One-dimensional spectra were corrected for atmospheric extinction and flux calibrated based on observations of flux standard stars \citep{Bohlin2010}.  At least one flux standard was observed on each night science data were obtained. An example of a flux-calibrated spectrum is shown in Figure \ref{fig:spectra}.   Given that each light path traversed different physical portions of the dichroic, we extract a 1D flat spectrum for each object and subsequently correct the flux calibration of each spectrum individually using the ratio of the local flat spectrum and the flat spectrum extracted at the position of the standard star.

While many H\ii\ regions exist in the central region of NGC\,5457, we chose to  observe a few of these regions in multiple masks.  This serves to confirm that our calibrations as stable and robust against changing observing conditions and permits us to obtain even deeper spectra in metal rich regions near the center of NGC\,5457 where the innate weakness of the auroral lines makes them difficult to detect.  For regions with coverage in multiple slit masks, in general, we coadd individual flux calibrated spectra to create the final spectrum.  In two cases we did not coadd all the spectra as some were clearly more noisy.  In all cases, analyses of the individual spectra are consistent with the summed spectrum.

\subsection{Spectral Modeling and Line Intensities}
We provide a detailed description of the adopted continuum modeling and line fitting procedures applied to the CHAOS observations in \citet{berg2015}.  Here we will only highlight the principle components of the process.  We model  the underlying stellar continuum using the STARLIGHT\footnote{www.starlight.ufsc.br} spectral synthesis code \citep{starlight} in conjunction with the models of \citet{bruzual}.  This allows us to fit for stellar absorption which is blended with the observed fine-structure lines.  

Allowing for an additional faint nebular continuum, modeled as a linear component, we fit Gaussian profiles to each emission line.  We note that blended emission lines in the Balmer sequence at the blue end of the spectrum  (H7, H8, and H11 -- H14) are not fit but are modeled based on the measurements of unblended Balmer lines and the tabulated atomic ratios of \citet{hummer}, assuming Case B recombination.  As the line profiles are not perfect Gaussian's, we subsequently use the modeled baselines and positions to integrate under the continuum subtracted spectrum, for all non-blended lines.  In general, the difference between the flux returned by a Gaussian fit and direct integration is less than 2\% of the line flux.  

Given that precise measurements of the auroral lines are vital to this study, we measured the flux of each auroral line by hand in the extracted spectra to confirm the fit.  While the majority of measurements were found to be in agreement, in cases where these measurements were in disagreement, we adopted the by hand measurement.  This was most common for the [N\,\ii] $\lambda$5755 line which falls in the wavelength region affected by the dichroic cutoff of MODS and the continuum here is not always well fit by the modeled starlight and the resultant fit was thus inadequate. 

We correct the strength of emission features for line-of-sight reddening using the relative intensities of the three strongest Balmer lines (H$\alpha$/H$\beta$, H$\alpha$/H$\gamma$, H$\beta$/H$\gamma$).  We report the determined values of c(H$\beta$) in Table \ref{t:lineflux} (the full table is available online).  We do not apply an ad-hoc correction to account for Balmer absorption as the lines were fit on top of stellar population models that accounted for this feature.  We note that the stellar models contain stellar absorption with an equivalent width of $\approx$1\,--\,2 \AA\ in the H$\beta$ line.  We report reddening-corrected line intensities measured from H\ii\ regions in the target fields in Table \ref{t:lineflux}.  

The uncertainty associated with each measurement is determined from measurements of the spectral variance, extracted from the two-dimensional variance image, Poisson noise in the continuum, read noise, sky noise, flat fielding calibration error, error in continuum placement, and error in the determination of the reddening.  We add, in quadrature, an additional 2\% uncertainty based on the precision of the adopted flux calibration standards \citep[][see discussion in Berg et al. 2015]{Bohlin2010}.   

\subsection{Diagnostic Diagrams}
We targeted H\ii\ regions using narrowband H$\alpha$ imaging which was continuum subtracted.  While H$\alpha$ is prominent in the ionized gas of an H\ii\ region, other objects besides H\ii\ regions can also produce such emission.  For example, planetary nebulae and supernovae remnants can be targeted based on the presence of ionized gas.  To verify that we are measuring lines from photoionized H\ii\ regions, we inspected standard diagnostic diagrams \citep[][BPT]{baldwin1981}: [O \iii]/H$\beta$ versus [N\,\ii]/H$\alpha$, [\ion{S}{2}]/H$\alpha$, and [\ion{O}{1}]/H$\alpha$, and ionization parameter P versus R$_{23}$, where
\begin{equation}
\rm{P} \equiv \frac{[\rm{O\,III}]\,\lambda\lambda\,4959,5007}{[\rm{O\,II}]\,\lambda\,3727 + [\rm{O\,III}]\,\lambda\lambda\,4959,5007}
\end{equation}
\begin{equation}
\rm{R}_{23}\,\equiv \frac{[\rm{O\,III}]\,\lambda\lambda\,4959,5007+ [O\,II]\,\lambda3727}{\rm{H}\beta}.
\end{equation}

The standard BPT diagram, [O \iii]/H$\beta$ versus [N\,\ii]/H$\alpha$, does not show any unusual regions.  However, there are five regions, NGC5457-250.8-52.0, NGC5457+44.7+153.7, NGC5457+650.1+270.7, NGC5457+299.1+464.0, and NGC5457-345.5+273.8, that stand out in both [\ion{S}{2}]/H$\alpha$ and [\ion{O}{1}]/H$\alpha$ relative to [O \iii]/H$\beta$. We have performed our analysis both including and excluding these data and found they alter neither the conclusions nor the fits to the abundance trends.  

We also note that three regions, NGC5457-12.3-271.1, NGC5457-219.4+308.7, and NGC5457-167.8+321.5, lie offset from the trend of ionization parameter with R$_{23}$ in the sense of a low ionization parameter for the given R23. However, all other line diagnostics for these three regions are normal, and the spectra do not stand out as unusual.  Therefore, we include these regions in our fits.

\section{Gas-Phase Abundances}
The ability to determine an elemental abundance from an emission spectrum is dependent on knowledge of (1) the electron density (n$_e$), (2) the electron temperature (T$_e$), and (3) a correction factor for unobserved ionic states.  Observationally, the challenge is to detect the intrinsically faint auroral lines which, when paired with their stronger ionic counterparts, are sensitive to electron temperature \citep[e.g., see][]{agn3}.  Lacking detections of these lines, one must turn to indirect methods, wherein measurements of the strong-lines have been calibrated either empirically or via photoionization modeling \citep[e.g.,][]{edmunds1984}.  In calculating temperatures and abundances we have adopted the updated atomic data presented in \citet{berg2015}. 

\subsection{Temperature Relations}
We have measured electron temperatures using five ratios of auroral to nebular lines: \oiii\,$\lambda$4363/$\lambda\lambda$4959,5007; \siii\,$\lambda$6312/$\lambda\lambda$9069, 9532; \nii\,$\lambda$5755/$\lambda\lambda$6548, 6583; \oii\,$\lambda\lambda$7319, 7330/$\lambda\lambda$3726, 3729; \sii\,$\lambda\lambda$4069, 4076/$\lambda\lambda$6717, 6731.  The derived electron temperatures from the different ions are reported in Table \ref{t:labundances}.  Different ions measure the electron temperature in physically different portions of an H\ii\ region, depending on the energy needed to excite the relevant transitions, and thus present independent temperature scales.  We have chosen to represent the temperature structure of an H\ii\ region with a three-zone model.  The different zones are characterized by different ionic temperatures. For the ionization stages represented in each zone we adopt: [O\,\ii], [N\,\ii], [S\,\ii] in the low-ionization zone; [S\,\iii], [Ar\,\iii] in the intermediate-ionization zone; and [O \iii], [Ne \iii] in the high-ionization zone.

One primary reason we selected NGC\,5457 for this study was the large number of bright H\ii\ regions that enhance the likelihood that multiple auroral lines will be detected in any given H\ii\ region.  Indeed, the ability to compare multiple temperature measures for a given H\ii\ region allows us to compare the different temperature scales.  Previous investigations of these relations has found very good agreement between the photoionization models of \citet{garnett1992}, namely,
\begin{equation}
\rm{T[S\,III]} = 0.83 \rm{T[O\,III]} + 0.17~(10^4\rm{K}),
\end{equation}
\begin{equation}
\rm{T[N\,II]} = \rm{T[O\,II]} = 0.70 \rm{T[O\,III]} + 0.3~(10^4\rm{K}),
\end{equation}
and the empirically measured temperatures from multiple ions \citep[e.g.,][]{kennicutt2003}.  In \citet{croxall2015} we found good agreement between T\siii\ and T\nii\ in H\ii\ regions in NGC\,5194.  However, no H\ii\ regions in that sample have measurements of T\oiii\ with which they may be compared.  In \citet{berg2015} we detected both the \oiii\ and \siii\ auroral lines in several H\ii\ regions.  These showed an offset relative to equation (3) and significant scatter, calling into question the rote acceptance of T\oiii\ as ground truth.  Following our practice in \citet{berg2015} we  adopt a theoretical ratio of $\lambda$9532/$\lambda$9069 = 2.51 to correct for possible contamination of the red end of the spectrum by atmospheric absorption.  

The sensitivity of MODS on LBT allows us to greatly increase the number of regions used in a comparison of electron temperatures, even while restricting ourselves to a single galaxy, particularly when the selected galaxy has numerous luminous H\ii\ regions as is the case with NGC\,5457.  As was noted in \citet{berg2015}, the current atomic data lead to significant changes in the temperatures derived from \siii\ and \oii.  Given this change in atomic data since the \citet{garnett1992} models were run, changes in the resultant temperature relations would not be unreasonable.  

\begin{figure}[tp] 
\epsscale{1.}
   \centering
   \plotone{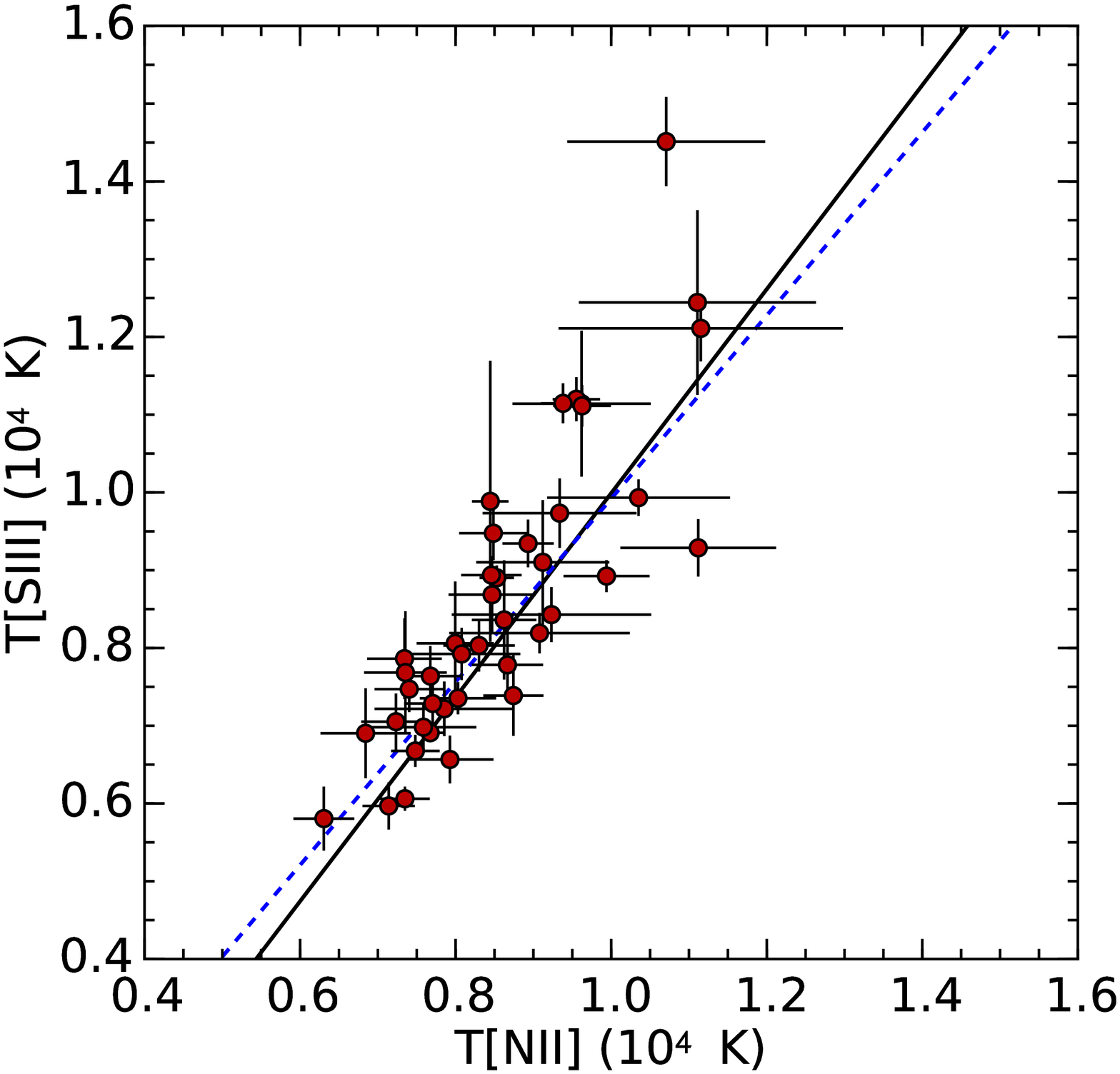}
   \plotone{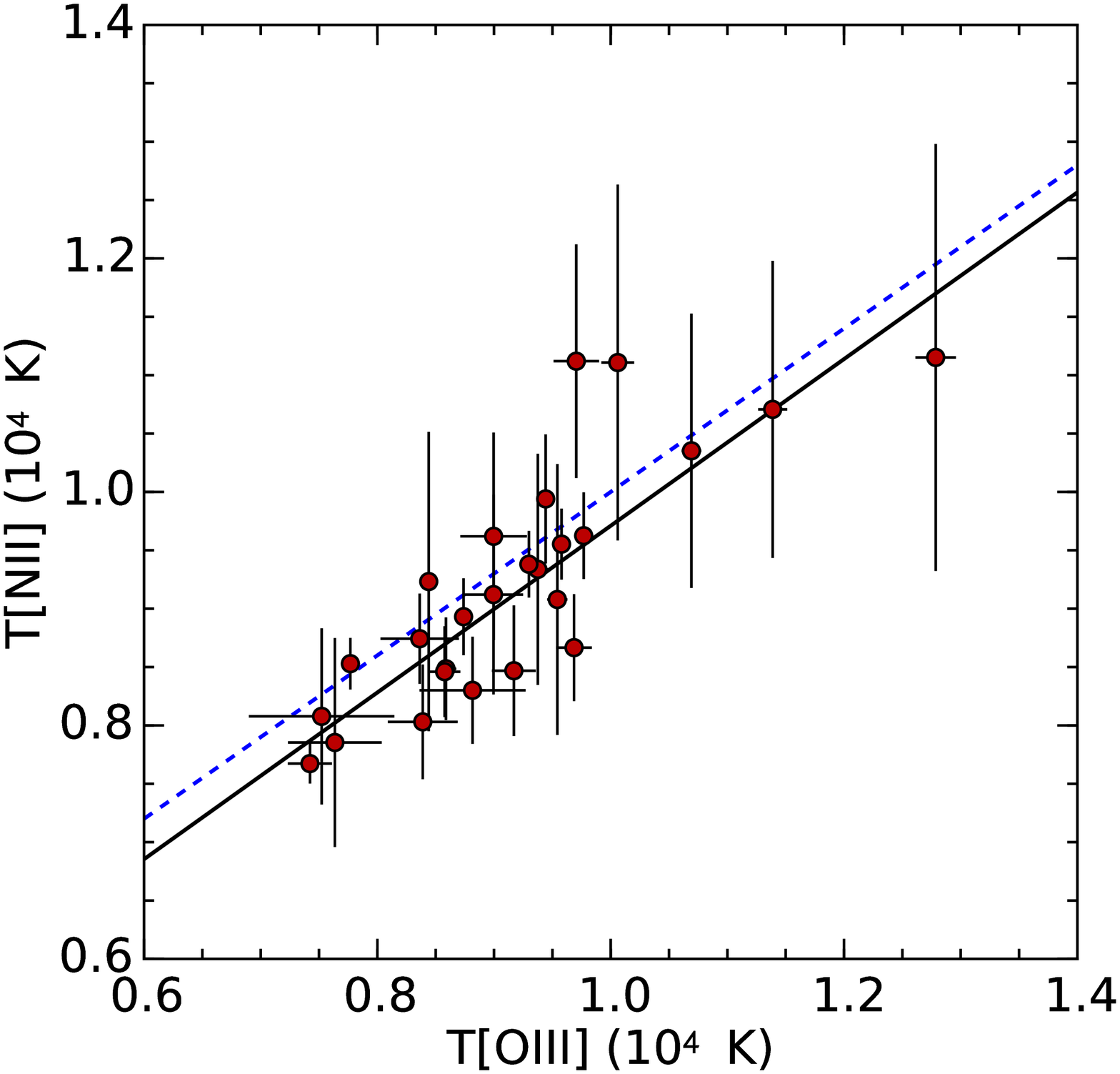}
   \plotone{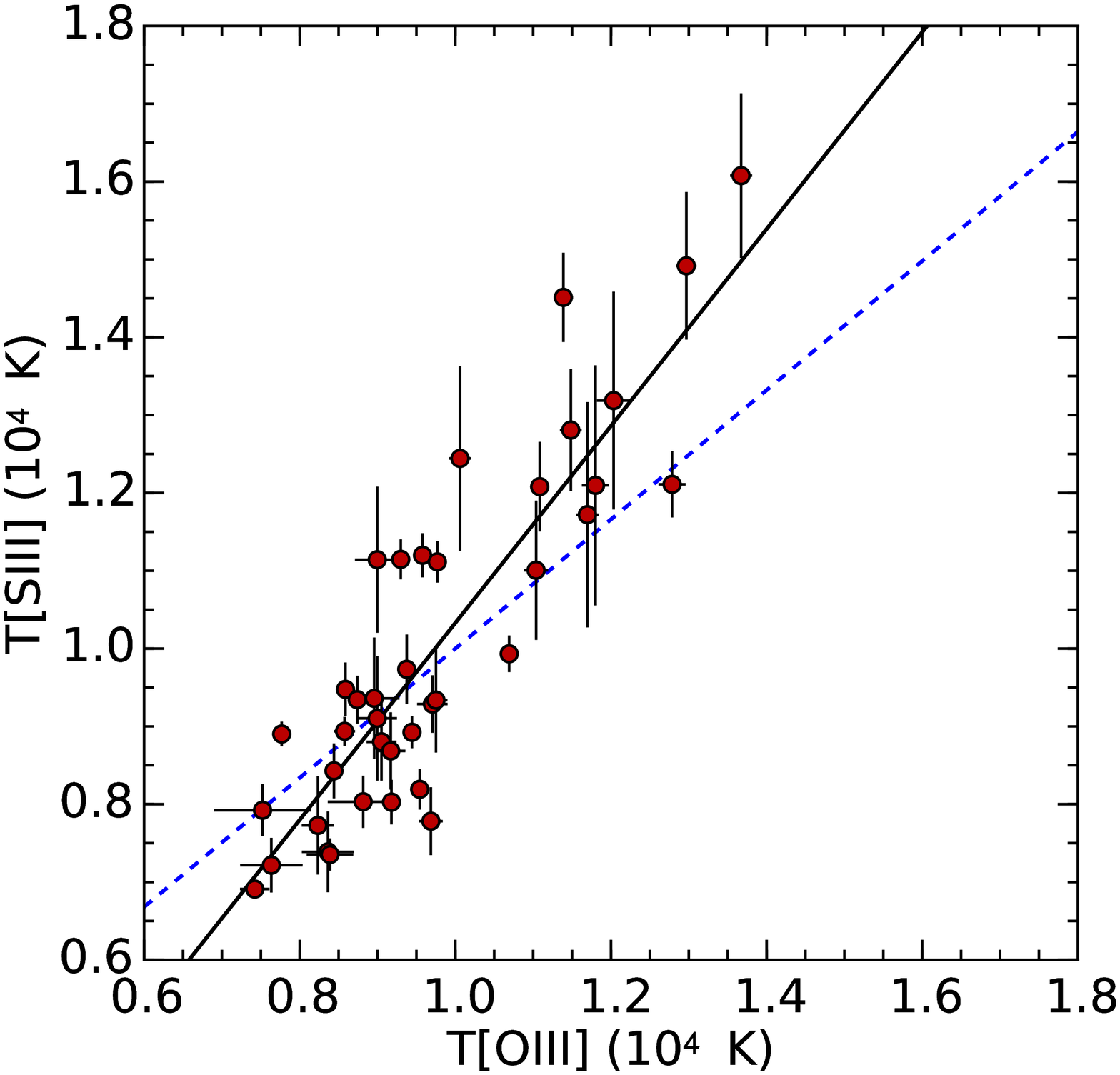}
    \caption{Relationship between T[N\,\ii] and T[S\,\iii] (Top), T[N\,\ii] (Middle), and  T[O\,\iii] and T[O\,\iii] and T[S\,\iii] (Bottom) measured for H\ii\ regions in NGC\,5457.  The empirical  relations from CHAOS observations as drawn as a solid black line while the photoionization models of \citet{garnett1992} are drawn as a dashed blue line. These temperature-temperature relations are characterized by intrinsic dispersions of 290$\pm$60\,K, 280$\pm$140\,K, and 1,050$\pm$160\,K }
   \label{fig:temperature}
\end{figure}

We plot the various temperature-temperature relations in Figure \ref{fig:temperature}.  Linear regression fits are computed using the Bayesian method described by \citet{kelly2007} that accounts for measurement errors in both variables and includes explicit fitting for intrinsic scatter in the regression \citep[see also][for a thorough discussion of this fitting problem in an astrophysical context]{hogg2010}.  This method models the intrinsic scatter in the independent variables as a linear mixture of Gaussian distributions.  We use J.\,MeyersÕ Python-language implementation of KellyÕs original IDL code\footnote{See github.com/jmeyers314/linmix} to perform the calculations.  This implementation uses a Markov chain Monte-Carlo to compute 10000 samplings from the posteriors.  We adopt as the ``best fit'' parameters the medians of the resulting marginal posterior distributions for the slope, intercept, intrinsic scatter, and linear correlation coefficients, and evaluate the standard deviation of these posterior distributions to estimate the uncertainties for each fit parameter.  We show the resulting fits as solid black lines; we also include equations (3) and (4), the relations derived from the photoionization models of \citet{garnett1992}, as dashed blue lines.  

As can be seen in the top panel of Figure \ref{fig:temperature}, we find a reasonable agreement with the models of \citet{garnett1992} when comparing T\nii\ and T\siii.  We do measure a slight shift in both the slope and zero-point offset in the empirical relationship, compared to the result of combining equations (3) and (4).  However, these differences are only at the one and two sigma level, respectively.  Furthermore, in the region where data exist, the two relations show a striking similarity.  These slight offsets are consistent with differences in atomic data we have adopted compared to the atomic data used by \citet{garnett1992}.

In the middle panel of Figure \ref{fig:temperature} we show the relationship between T\nii\ and T\oiii.  In this case we find agreement with the slope derived by \citet{garnett1992} and only note a slight shift in the zero-point offset of the relationship.  While the zero-point offset is just over the three sigma confidence level it can be explained given the differences in the adopted atomic data.  

Similar to the results of \citet{kennicutt2003}, we find a good correlation between T\oiii\ and T\siii\ in the 37 regions where both temperature sensitive line ratios are measured, as shown in the bottom panel of Figure \ref{fig:temperature}.  However, in this case, we do not find good agreement with the models of \citet{garnett1992}. Again, this may, in part, also arise from the different atomic data adopted, however, the atomic data alone cannot explain the difference.  Rather, as shown in Figure \ref{fig:temperature} the empirical relationship between these two ionic temperatures is significantly steeper than that given by equation (3).  This is particularly important for cool H\ii\ regions where the \siii\ auroral line is more frequently detected relative to the \oiii\ auroral line.  Indeed, in the coolest regions where the \siii\,$\lambda$6312 line is detected, the two different relations lead to temperatures that differ by more than 2,000\,K.  We also note that, in contrast with what was seen in NGC\,628 \citep{berg2015}, the relationship between T\oiii\ and T\siii\ is fairly tight with a root-mean-square-deviation of $\approx$2200\,K, compared to the scatter we measure in the NGC\,0628 data, $\approx$5000\,K.  Nevertheless, we do note that there are H\ii\ regions which lie below this relation in the regime largely populated in NGC\,628.

Two H\ii\ regions, NGC5457-397.4-71.7 and NGC5457-226.9-366.4, exhibit anomalously high T\nii\ and have been excluded from these fits and plots. \citet{kennicutt2003} also noted some discrepant points when comparing these two temperatures and concluded that unphysically high T\nii\ were likely the result of marginally detected lines.  Careful investigation of our spectra do not support a similar conclusion for our two discrepant data points.  Indeed, in the most discrepant of the two regions, NGC5457-397.4-71.7, the \nii\,$\lambda$5755 line can be seen in each of the six individual two dimensional images, ensuring it is not an anomalous measurement due to contamination via cosmic rays. Rather these anomalous temperatures likely indicate a complex environment in these giant star-forming complexes.   The existence of such unusual temperature signatures can lead to highly skewed results if only a single temperature measure is observed for a given H\ii\ region.  

Given that we derive abundances using different atomic data from those used in the models of \citet{garnett1992}, applying those temperature relations would introduce systematic offsets.  While previous data sets \citep[e.g.,][]{berg2015,croxall2015} did not cover a large enough range of parameter space to truly constrain empirical temperature-temperature relations, the number of observed H\ii\ regions with multiple temperature measurements and the steep oxygen gradient in NGC\,5457 \citep{kennicutt2003} result in the ideal uniform data set to derive empirical temperature relations.  We thus adopt the empirical temperature relations fit to CHAOS data for which multiple auroral lines were detected in a single region, namely:
\begin{equation}
T_{\rm{[N\,II]}} = 0.714\,(\pm0.142) T_{\rm{[O\,III]}} + 0.257\,(\pm0.125)~(10^4\rm{K}),
\end{equation}
 \begin{equation}
T_{\rm{[S\,III]}} = 1.312\,(\pm0.075) T_{\rm{[N\,II]}} - 0.313\,(\pm0.058)~(10^4\rm{K}),
\end{equation}
and
 \begin{equation}
T_{\rm{[S\,III]}} = 1.265\,(\pm0.140) T_{\rm{[O\,III]}} - 0.232\,(\pm0.135)~(10^4\rm{K}).
\end{equation}
These empirical relations are consistent with intrinsic dispersion of 280$\pm$140\,K, 290$\pm$60\,K, and 1,050$\pm$160\,K respectively.  We adopt these values as lower limits for the uncertainty of a temperature derived via these relations. To ensure that we are not introducing additional bias in our abundance determinations by adopting these temperature relations we confirm that offsets from equations (5) -- (7) do not correlate with the oxygen ion fraction, O$^+$/O, or excitation.   

In addition to measuring electron temperatures in each ionization zone of our three-zone model, we also measure three independent ionic ratios that characterize the low-ionization zone, T\oii, T\sii, and T\nii.  The standard practice is to adopt either one of the temperature-temperature relations already discussed or to measure T\nii.  This has arisen due to a large amount of scatter in T\oii\ and the relative weakness of the \sii\ auroral lines.  The increased dispersion in T\oii, compared to other measurements of electron temperature, has also been attributed to contributions from direct dielectric recombinations to the $^2P^0$ level, from which the auroral lines arise \citep{rubin1986}.  Furthermore, in the case of T\oii\, underlying telluric absorption, proximity of strong OH Meinel band emission, possible contribution due to recombination \citep{Liu2000}, and a susceptibility of \oii\ to collisional de-excitation complicate temperature measurements from \oii\ lines.  Additionally the temperature sensitive line ratios for \oii\ and \sii\ are separated by $\approx$3,600\,\AA\ and  $\approx$2,600\,\AA, respectively, making them more sensitive to the effects of reddening.  
 \begin{figure}[tbp] 
\epsscale{1.}
   \centering
   \plotone{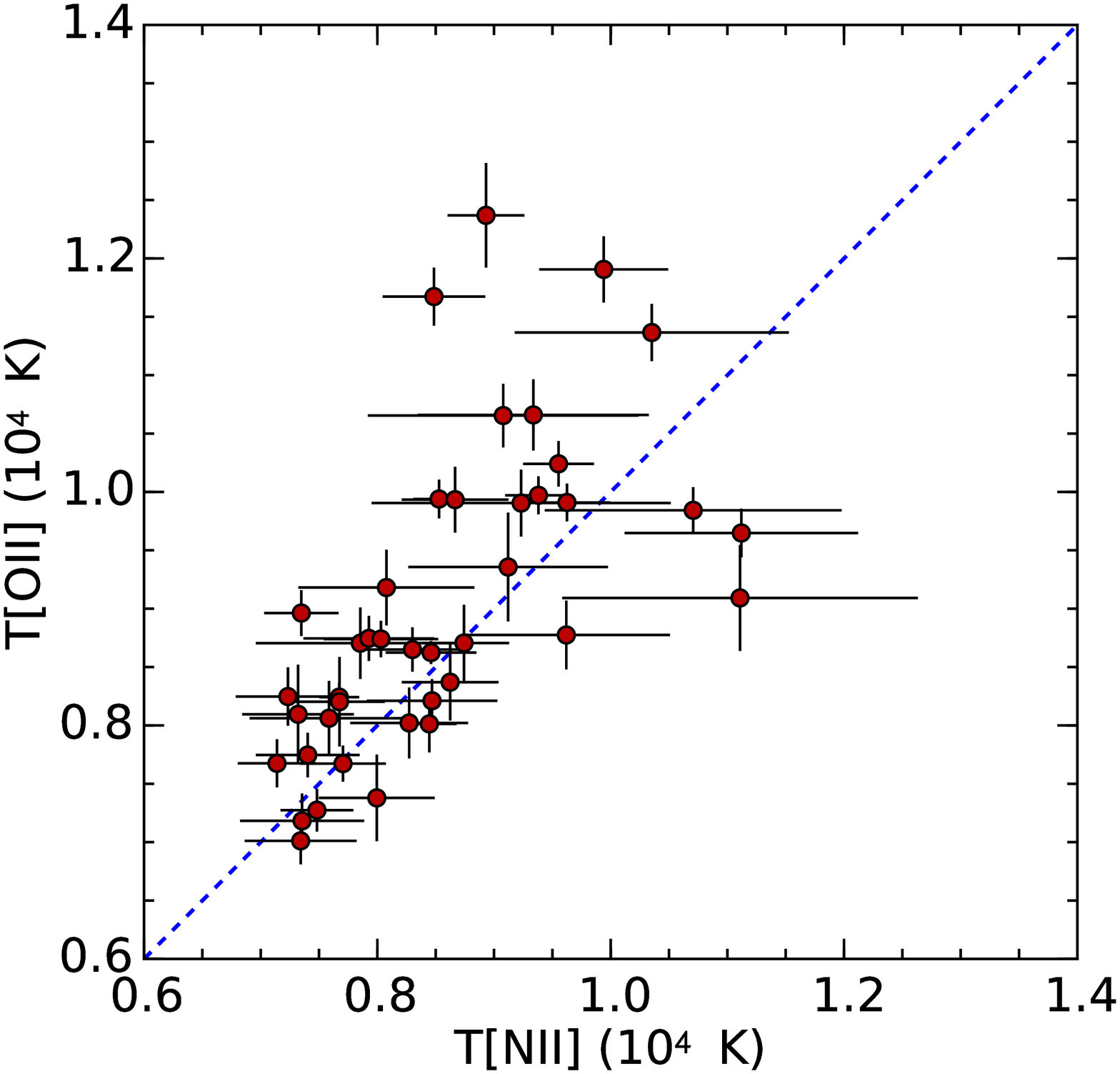}\\
   \plotone{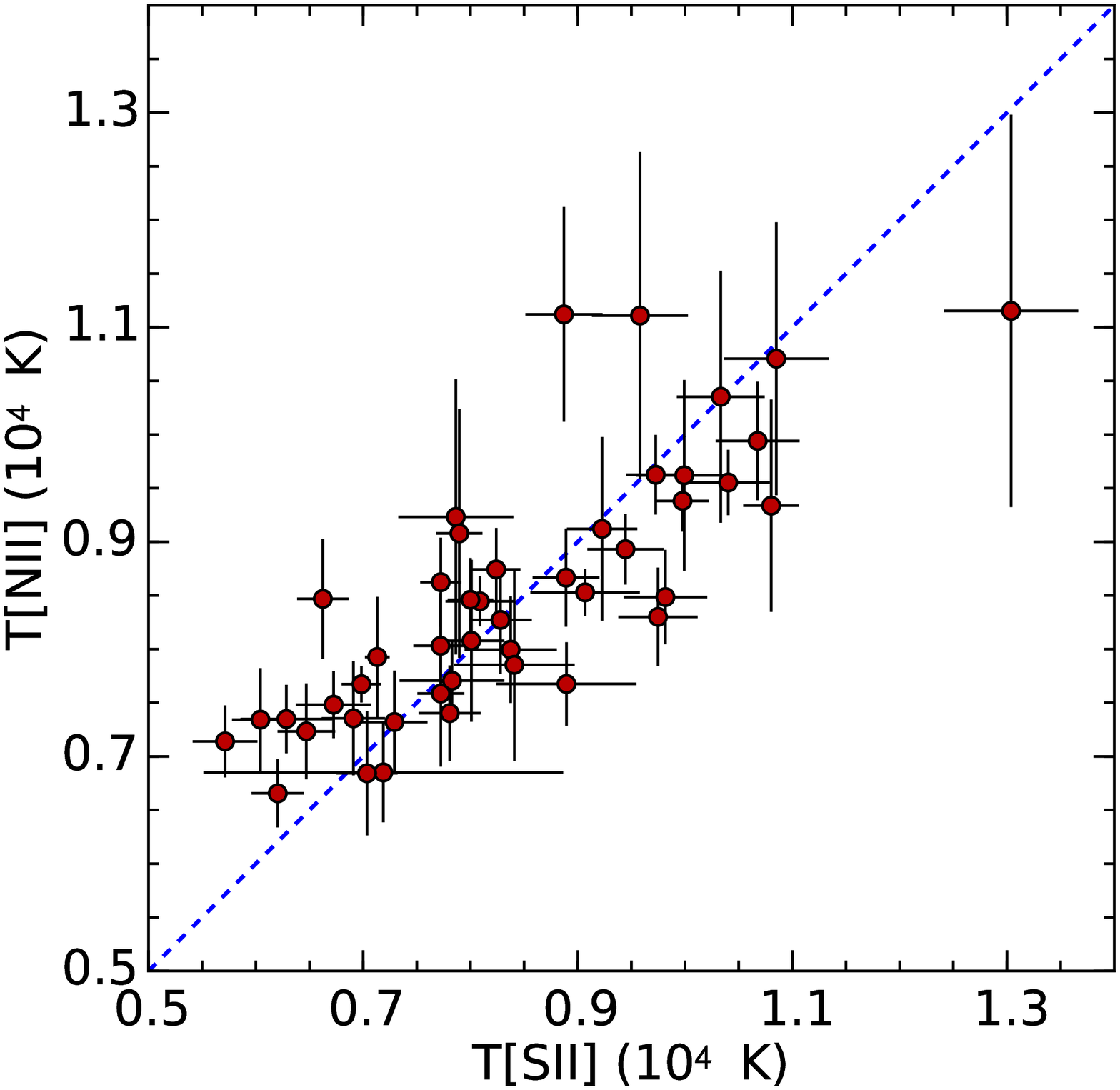}
    \caption{Relationship between T[N\,\ii] and T[O\,\ii] (Top) and T[N\,\ii] and T[S\,\ii] (Bottom) measured for H\ii\ regions in NGC\,5457.  The assumed one-to-one line is plotted showing that the data are in agreement with these ions being isothermal.  }  
   \label{fig:temperature2}
\end{figure}  

Our measurements of T\oii, T\sii, and T\nii\ do indeed suggest that they largely arise from a common ionization zone.  In Figure \ref{fig:temperature2} we plot the measured T\oii\ (Top) and  T\sii\ (Bottom) against the measured T\nii for H\ii\ regions where both respective sets of auroral lines were measured.  While there is noticeable scatter about the one-to-one relation, with intrinsic dispersions of $\approx$900\,K for T\sii\ and $\approx$1,100\,K for T\oii, the three temperatures are clearly correlated and suggest that the three ions are roughly measuring the electron temperature in an equivalent volume of gas.  While an exhaustive investigation of the \oii\ and \sii\ lines is not the purpose of this paper, we have looked for possible trends in the deviations of T\oii\ and T\sii\ from T\nii.  We do not find any significant correlations with a variety of parameters such as the ionization parameters and reddening coefficient that might be expected given previous explanations of this scatter.  Notably, restricting the sample to only the highest signal to noise detections does not decrease the scatter between these three temperatures.  

\subsection{Comparison with Previous Work}
Electron temperatures have previously been reported in a significant number of H\ii\ regions in NGC\,5457, making it a useful benchmark test for the CHAOS program.  Notably, we have significant overlap with the samples of \citet{kennicutt2003} (14/20 regions), \citet{bresolin2007} (2/2 regions), and \citet{li2013} (6/10 regions).  Given that we adopt slightly different atomic data \citep{berg2015}, we take their reported line measurements and re-compute electron temperatures and abundances.   

\begin{figure}[tbp] 
\epsscale{1.0}
   \centering
    \plotone{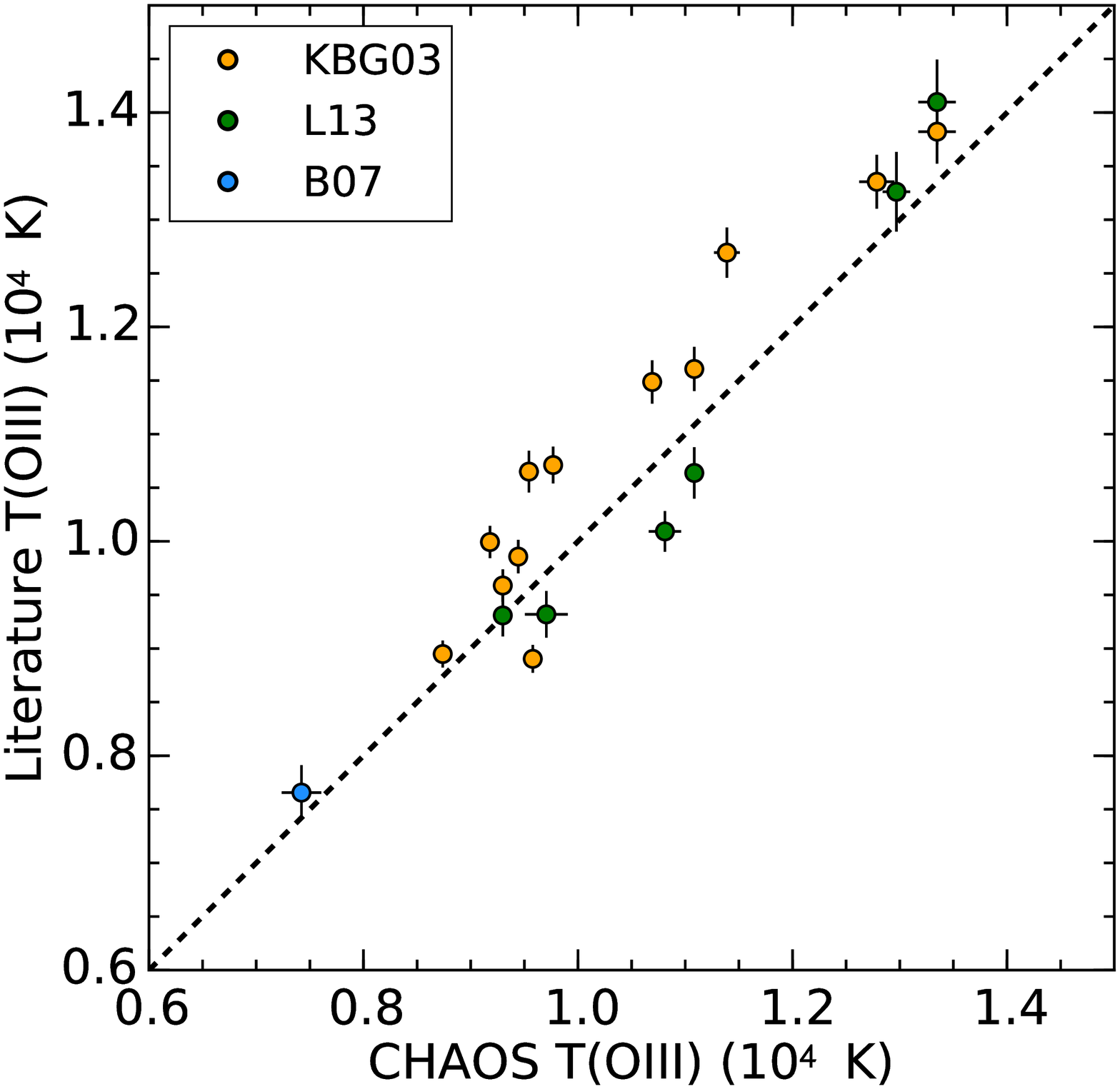}\\
    \plotone{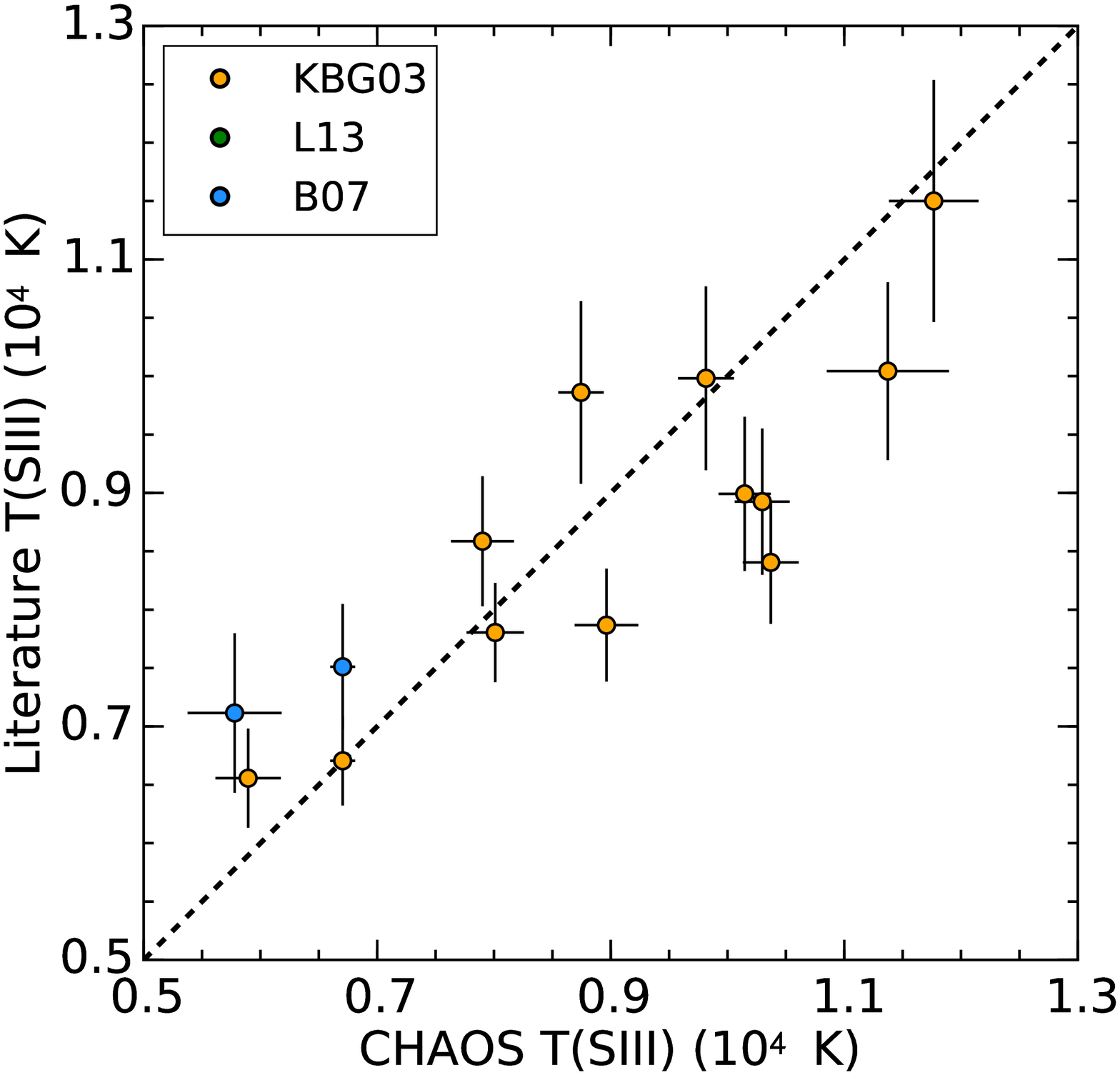}\\
    \plotone{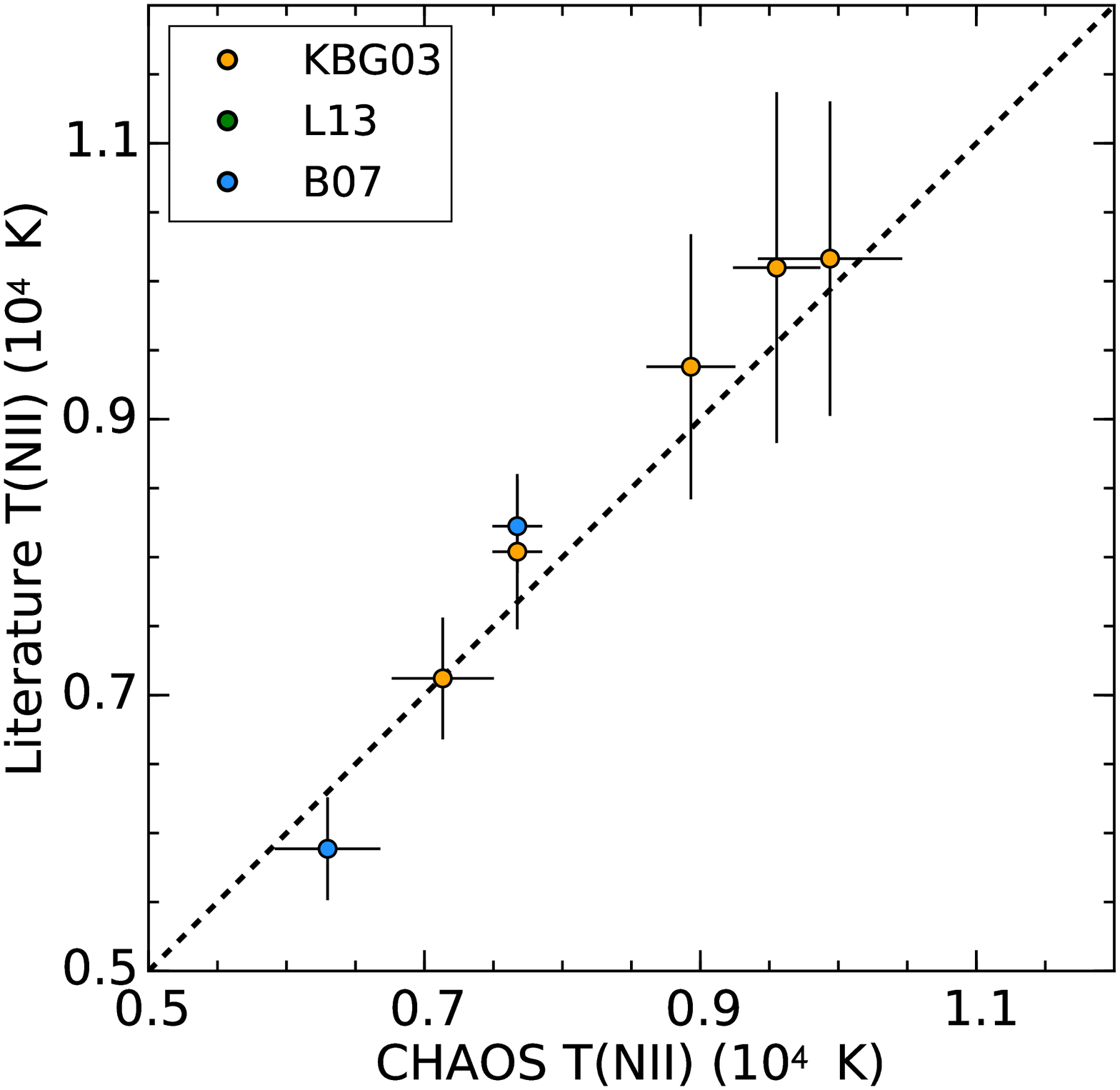}
   \caption{Comparison between temperatures derived from CHAOS spectra and those derived from the literature. A one-to-one correlation is plotted as a dashed gray line. }  
   \label{fig:temperature_comp}
\end{figure}

In Figure \ref{fig:temperature_comp}, we compare the derived electron temperatures from the line measurements reported in \citet{kennicutt2003}, \citet{bresolin2007}, and \citet{li2013} with those reported here.  As can be seen, the calculated temperatures lie close to a one-to-one relation for all three ions used to independently derive the temperature.  In warm regions, the \nii\,$\lambda$5755 line becomes more difficult to detect.  This is reflected in the larger errors in the warmer H\ii\ regions.  Indeed, in two regions, NGC5457-371.1-280.0 (H143) and NGC5457-368.3-285.6 (H149),  \citet{kennicutt2003} adopt significantly lower temperatures for the low ionization region based on temperatures derived from the \oiii\ measurements.  We do not plot these two regions in the right panel of Figure \ref{fig:temperature_comp}; however, we note that while their measured T\nii\ is highly discrepant from ours, their adopted value is in agreement with our measurements.

In addition to electron temperatures, we report line strengths and the recomputed oxygen, nitrogen, and sulfur abundances for H\ii\ regions measured in the literature and our newly obtained data in Table \ref{t:comptable}.  In general, we find very good agreement between the abundances derived from reported line strengths in \citet{kennicutt2003}, \citet{bresolin2007}, and \citet{li2013} and those presented in this work, with an average difference in 12+log(O/H) and log(N/O) of 0.02 dex and 0.08 respectively.  

While most H\ii\ regions we have in common with previous reported studies are in agreement within the stated errors, two regions exhibit significant deviations.  Comparing abundance determinations in NGC\,5457+354.1+71.2 and NGC\,5457-75.0+29.3 reveals that our new line measurements yield gas-phase oxygen abundances that are discrepant at the level of 0.25\,dex, well above the stated errors, compared to recomputed abundances.   However, the recomputed abundances are also not in agreement with the original computed abundances.  In the case of NGC\,5457+354.1+71.2, \citet{kennicutt2003} do not adopt the temperature derived from the \oiii\,$\lambda$4363 line.  Rather, they adopt a significantly lower temperature for this H\ii\ region, which places it in agreement with our CHAOS observations.  On the other hand the individual electron temperatures derived for NGC\,5457-75.0+29.3 are all in agreement, while specific ionic abundances seem to be highly discrepant.  The large differences for the recomputed abundances in these two regions could easily be due to typographical errors in the published tables of line strengths, so we compare our CHAOS data with the published abundances from the original works.

\begin{deluxetable*}{lccccccccccc}  
\tabletypesize{\scriptsize}
\tablecaption{Comparison with Previous Observations}
\tablewidth{0pt}
\tablehead{ 
  \colhead{H\ii\ Region}	&
  \colhead{$\frac{\rm{R}}{\rm{R}_{25}}$}	&
  \colhead{[O\,\iii]\,$\lambda$5007}	&
  \colhead{[O\,\iii]\,$\lambda$4363}	&
  \colhead{[N\,\ii]\,$\lambda$5755}	&
  \colhead{[S\,\iii]\,$\lambda$6312}	&
  \colhead{ T[O\,\iii]}	&
  \colhead{ T[N\,\ii] }	&
  \colhead{ T[S\,\iii]}	&
  \colhead{12+log($\frac{\rm{O}}{\rm{H}}$)}	&
  \colhead{log($\frac{\rm{N}}{\rm{O}}$)}	&
  \colhead{log($\frac{\rm{S}}{\rm{O}}$)}	
  }
\startdata
-75.0+29.3	&	0.10	&	6	&	\ldots	&	0.32	&	0.22	&	\ldots	&	6306	&	5806	&	8.56	&	-0.58	&	-1.26	\\\vspace{0.06in}
H493 (B07)$^\dagger$	&	0.10	&	8	&	\ldots	&	0.24	&	0.27	&	\ldots	&	5960	&	7330	&	8.74	&	0.72	&	-1.78	\\
																							
+164.6+9.9	&	0.20	&	98	&	0.20	&	0.44	&	0.79	&	7422	&	7673	&	6910	&	8.57	&	-0.96	&	-1.23	\\
H1013 (K03)	&	0.19	&	103	&	\ldots	&	0.50	&	0.80	&	\ldots	&	8040	&	6705	&	8.47	&	-0.90	&	-1.04	\\\vspace{0.06in}
H1013 (B07)	&	0.19	&	102	&	0.24	&	0.59	&	0.77	&	7656	&	8225	&	7512	&	8.48	&	-0.91	&	-1.34	\\
																							
-159.9+89.6	&	0.22	&	19	&	\ldots	&	0.51	&	0.35	&	\ldots	&	7140	&	5969	&	8.55	&	0.77	&	-1.22	\\\vspace{0.06in}
H336 (K03)	&	0.22	&	23	&	\ldots	&	0.50	&	0.60	&	\ldots	&	7121	&	6557	&	8.61	&	-0.82	&	-1.29	\\
																							
+254.6-107.2	&	0.33	&	346	&	1.39	&	0.40	&	1.54	&	8739	&	8931	&	9344	&	8.48	&	-1.00	&	-1.58	\\\vspace{0.06in}
H1105 (K03)	&	0.34	&	316	&	1.40	&	0.40	&	1.30	&	8949	&	9380	&	7867	&	8.42	&	-1.08	&	-1.27	\\
																							
+354.1+71.2	&	0.43	&	352	&	1.94	&	0.37	&	1.40	&	9542	&	9079	&	8192	&	8.45	&	-1.20	&	-1.17	\\\vspace{0.06in}
H1170 (K03)$^\dagger$	&	0.42	&	201	&	1.60	&	\ldots	&	1.70	&	10800	&	\ldots	&	8500	&	8.37	&	1.20	&	1.44	\\
																							
+360.9+75.3	&	0.44	&	354	&	1.71	&	\ldots	&	1.28	&	9179	&	\ldots	&	8027	&	8.43	&	-1.18	&	-1.34	\\\vspace{0.06in}
H1176 (K03)	&	0.43	&	369	&	2.40	&	\ldots	&	1.50	&	9993	&	\ldots	&	8586	&	8.30	&	-1.16	&	-1.27	\\
																							
-99.6-388.0	&	0.47	&	388	&	2.08	&	0.36	&	1.41	&	9443	&	9940	&	8925	&	8.38	&	-1.15	&	-1.47	\\\vspace{0.06in}
H409 (K03)	&	0.48	&	370	&	2.30	&	0.40	&	1.70	&	9858	&	10160	&	9861	&	8.33	&	-1.17	&	-1.51	\\
																							
-226.9-366.4	&	0.50	&	151	&	0.89	&	0.73	&	1.28	&	9705	&	11120	&	9288	&	8.18	&	-1.18	&	-1.27	\\\vspace{0.06in}
H219 (Li13)	&	0.50	&	178	&	0.91	&	\ldots	&	\ldots	&	9319	&	\ldots	&	\ldots	&	8.51	&	-1.43	&	\ldots	\\
																							
-368.3-285.6	&	0.54	&	323	&	1.64	&	0.43	&	1.40	&	9299	&	9381	&	11150	&	8.42	&	-1.13	&	-1.54	\\
H149 (Li13)	&	0.54	&	329	&	1.68	&	\ldots	&	\ldots	&	9309	&	\ldots	&	\ldots	&	8.46	&	-1.20	&	\ldots	\\\vspace{0.06in}
H149 (K03)	&	0.55	&	318	&	1.80	&	0.90	&	1.40	&	9588	&	13050	&	8925	&	8.37	&	-1.08	&	-1.17	\\
																							
-371.1-280.0	&	0.54	&	267	&	1.61	&	0.48	&	1.28	&	9768	&	9625	&	11110	&	8.33	&	-1.13	&	-1.52	\\\vspace{0.06in}
H143 (K03)	&	0.55	&	284	&	2.30	&	0.90	&	1.40	&	10711	&	13600	&	8992	&	8.18	&	-1.06	&	-1.05	\\
																							
-455.7-55.8	&	0.55	&	325	&	2.92	&	\ldots	&	1.28	&	11080	&	\ldots	&	12080	&	8.22	&	-1.43	&	-1.61	\\\vspace{0.06in}
H59 (Li13)	&	0.55	&	290	&	2.30	&	\ldots	&	\ldots	&	10640	&	\ldots	&	\ldots	&	8.30	&	-1.54	&	\ldots	\\
																							
-392.0-270.1	&	0.55	&	372	&	2.10	&	0.29	&	1.31	&	9578	&	9553	&	11200	&	8.35	&	-1.14	&	-1.80	\\\vspace{0.06in}
H128 (K03)	&	0.56	&	391	&	1.70	&	0.30	&	1.30	&	8904	&	10100	&	8405	&	8.42	&	-1.10	&	-1.32	\\
																							
-455.7-55.8	&	0.55	&	325	&	2.92	&	\ldots	&	1.28	&	11080	&	\ldots	&	12080	&	8.22	&	-1.43	&	-1.61	\\\vspace{0.06in}
H67 (K03)	&	0.56	&	342	&	3.50	&	\ldots	&	1.80	&	11610	&	\ldots	&	10040	&	8.14	&	-1.38	&	-1.12	\\
																							
-453.8-191.8	&	0.58	&	374	&	3.64	&	0.32	&	1.44	&	11390	&	10710	&	14510	&	8.21	&	-1.37	&	-1.67	\\\vspace{0.06in}
H71 (K03)	&	0.57	&	454	&	5.90	&	\ldots	&	1.90	&	12690	&	\ldots	&	\ldots	&	8.06	&	-1.26	&	\ldots	\\
																							
-481.4-0.5	&	0.58	&	284	&	3.92	&	\ldots	&	1.31	&	12970	&	\ldots	&	14920	&	8.02	&	-1.37	&	-1.61	\\\vspace{0.06in}
H27 (Li13)	&	0.58	&	251	&	3.64	&	\ldots	&	\ldots	&	13260	&	\ldots	&	\ldots	&	8.01	&	-1.39	&	\ldots	\\
																							
+315.3+434.4	&	0.62	&	441	&	3.67	&	\ldots	&	\ldots	&	10810	&	\ldots	&	\ldots	&	8.32	&	-1.31	&	-1.40	\\\vspace{0.06in}
H1146 (Li13)	&	0.63	&	457	&	3.07	&	\ldots	&	\ldots	&	10090	&	\ldots	&	\ldots	&	8.40	&	-1.40	&	\ldots	\\
																							
+509.5+264.1	&	0.67	&	417	&	3.37	&	0.25	&	1.53	&	10690	&	10350	&	9933	&	8.26	&	-1.33	&	-1.52	\\\vspace{0.06in}
H1216 (K03)	&	0.67	&	473	&	4.70	&	\ldots	&	1.60	&	11490	&	\ldots	&	9982	&	8.17	&	-1.34	&	-1.40	\\
																							
+667.9+174.1	&	0.81	&	674	&	8.85	&	0.16	&	1.64	&	12790	&	11150	&	12110	&	8.14	&	-1.37	&	-1.65	\\\vspace{0.06in}
N5471-A (K03)	&	0.81	&	644	&	9.50	&	\ldots	&	1.60	&	13354	&	\ldots	&	11500	&	8.05	&	-1.35	&	-1.43	\\
																							
+1.0+885.8	&	1.05	&	184	&	2.72	&	\ldots	&	\ldots	&	13350	&	\ldots	&	\ldots	&	7.85	&	-1.47	&	-1.59	\\
H681a (K03)	&	1.04	&	312	&	5.00	&	\ldots	&	\ldots	&	13820	&	\ldots	&	\ldots	&	7.91	&	-1.51	&	-1.55	\\
H681 (Li13)	&	1.05	&	260	&	4.36	&	\ldots	&	\ldots	&	14100	&	\ldots	&	\ldots	&	7.85	&	-1.55	&	\ldots			
\enddata
\label{t:comptable}
\tablecomments{H\ii\ regions that are in common between our new observations and \citet{kennicutt2003}, \citet{bresolin2007}, and \citet{li2013}.  For each H\ii\ region we give the offset from the galaxy center given in Table \ref{t:n5457global} on the first line along with the derived results from this study.  In the subsequent lines for each given region we report the name used in the comparison study as well as give an abbreviated name for each study, namely, K03 for \citet{kennicutt2003}, B07 for \citet{bresolin2007}, and L13 for \citet{li2013}.  Fluxes are given in units of H$\beta$\,=\,100.}
\tablenote[0]{$^\dagger$ These regions are compared to the published values from  \citet{kennicutt2003}, \citet{bresolin2007} as discussed in the text.}
\end{deluxetable*}

\subsection{Electron Densities and Temperature Selections}

To calculate abundances in regions where one or more temperature sensitive line ratios were measured, we adopt an electron density, n$_e$, based on the \sii\,$\lambda$6717/6731 line ratio as these relatively strong lines are cleanly separated in MODS1 spectra.   The large majority of H\ii\ regions observed, 95 out of 109, lie in the low-density regime [I($\lambda$6717)/I($\lambda$6731) $>$ 1.35].  For these regions we thus adopt an n$_e$ of 100 cm$^{-3}$.  For the remaining 14 regions we calculate electron densities using a five level atom code based on FIVEL \citep{fivel}, see Table \ref{t:labundances}.  Although these regions have densities that are higher than the low density limit, they are all still low enough that collisional de-excitation is not an important factor.  We note that the [O\,\ii] $\lambda\lambda$3726,3729 doublet is blended for all observations; however, in the majority of spectra, the doublet profile is clearly non-Gaussian.  We have modeled this doublet using two Gaussian profiles, for use primarily as a consistency check of the [S\,\ii] density determination.  For all calculations aside from this density check, we sum the flux in the [O\,\ii] $\lambda\lambda$3726, 3729 doublet.

In regions where we have measured  T\oiii, T\siii, and T\nii, we adopt the measured temperatures for each individual ionization zone when calculating abundances.  When one or more of these temperatures have not been measured, we use equations (5) -- (7) to calculate a temperature for the ionization zone that is not measured.  Given the scatter measured in the new temperature-temperature relations, where possible, T\nii\ is selected over T\siii\ where estimating T\oiii; similarly, T\oiii\ is preferentially selected to estimate T\nii.  When an ionic temperature is calculated via equations (5) -- (7), the intrinsic dispersion for the relation used is adopted as the uncertainty in that adopted temperature.  Both measured temperatures and the adopted temperatures are reported in Table \ref{t:labundances} (the full table is available online).  

\subsection{Ionization Correction Factors}
Nebular oxygen abundances obtained from optical spectra are quite straightforward.  Given the similar ionization potentials of neutral hydrogen and oxygen, 13.60\,eV and 13.62\,eV respectively, we can reasonably assume that all the oxygen in an H\ii\ region is ionized.  Furthermore, the high ionization potential of O$^{++}$, 54.94\,eV, means that O$^{+++}$ is also negligible in typical H\ii\ regions.  Thus, we can account for all oxygen ions present by summing the number of singly and doubly ionized oxygen ions.  Unfortunately, other elements do not have a correspondingly simple pairing of observable lines in all ionization species and we must account for the presence of unobserved ionization states.

While many studies have investigated ionization correction factors, most empirical studies are focused on higher ionization nebulae where auroral lines are easier to detect.  By expanding our study towards low ionization H\ii\ regions, we must explore different options in ionization correction factors and the regime in which each correction is valid.   We refer the interested reader to Appendix A wherein we discuss the details of different ionization correction factors for each element of interest to this work.  

To derive nitrogen abundances we assume N/O = N$^+$/O$^+$, given the well matched ionization potential of these two species.  Neon abundances were derived under the assumption Ne/O = Ne$^{++}$/O$^{++}$ \citep{peimbert1969}.   The ionization correction factor for sulfur becomes more challenging in low ionization nebulae as the parameter space has not been fully explored empirically.  Thus, we adopt  
\begin{equation}
\frac{S}{O} = \frac{S^+ + S^{++}}{O^+},
 \end{equation} 
for O$^+$/O $\geq$0.4 and the ionization correction of \citet{thuan1995} for O$^+$/O $\leq$0.4.  Similar to  \citet{kennicutt2003} we find the ratio of Ar$^{++}$/S$^{++}$ traces the Ar/S ratio.  However for low ionization nebulae we adopt a linear correction to Ar$^{++}$/S$^{++}$,
 \begin{equation}
\rm{log}\frac{Ar^{++}}{S^{++}} = -1.049\frac{O^+}{O} -0.022 \mbox{, for } \frac{O^+}{O}  \geq 0.6,
 \end{equation}
to account for an increase in the Ar$^+$ population.   We report all ionization correction factors and abundance ratios in Table \ref{t:labundances}.

\subsection{Abundance gradients}
\begin{figure*}[tbp] 
\epsscale{0.55}
   \centering
   \plotone{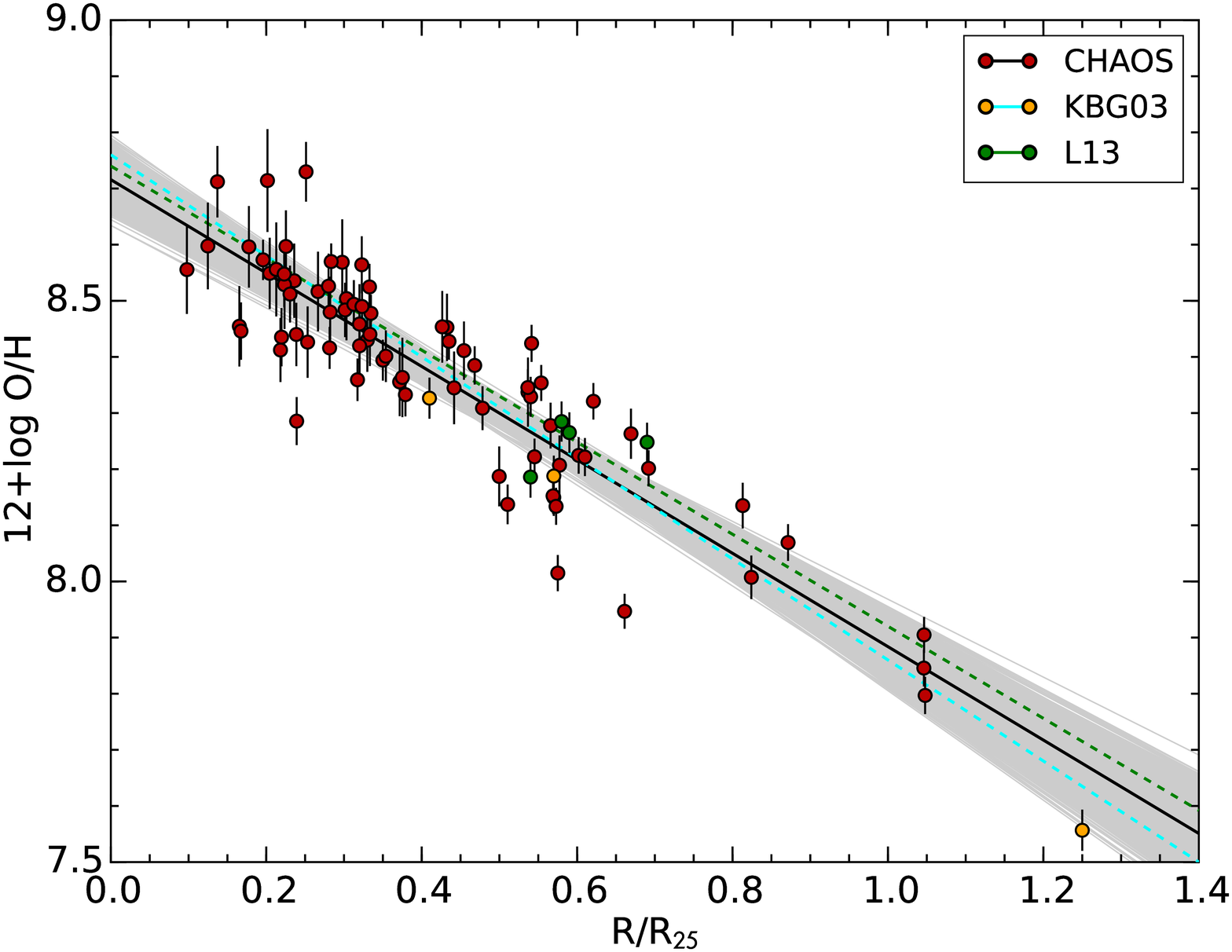}
   \plotone{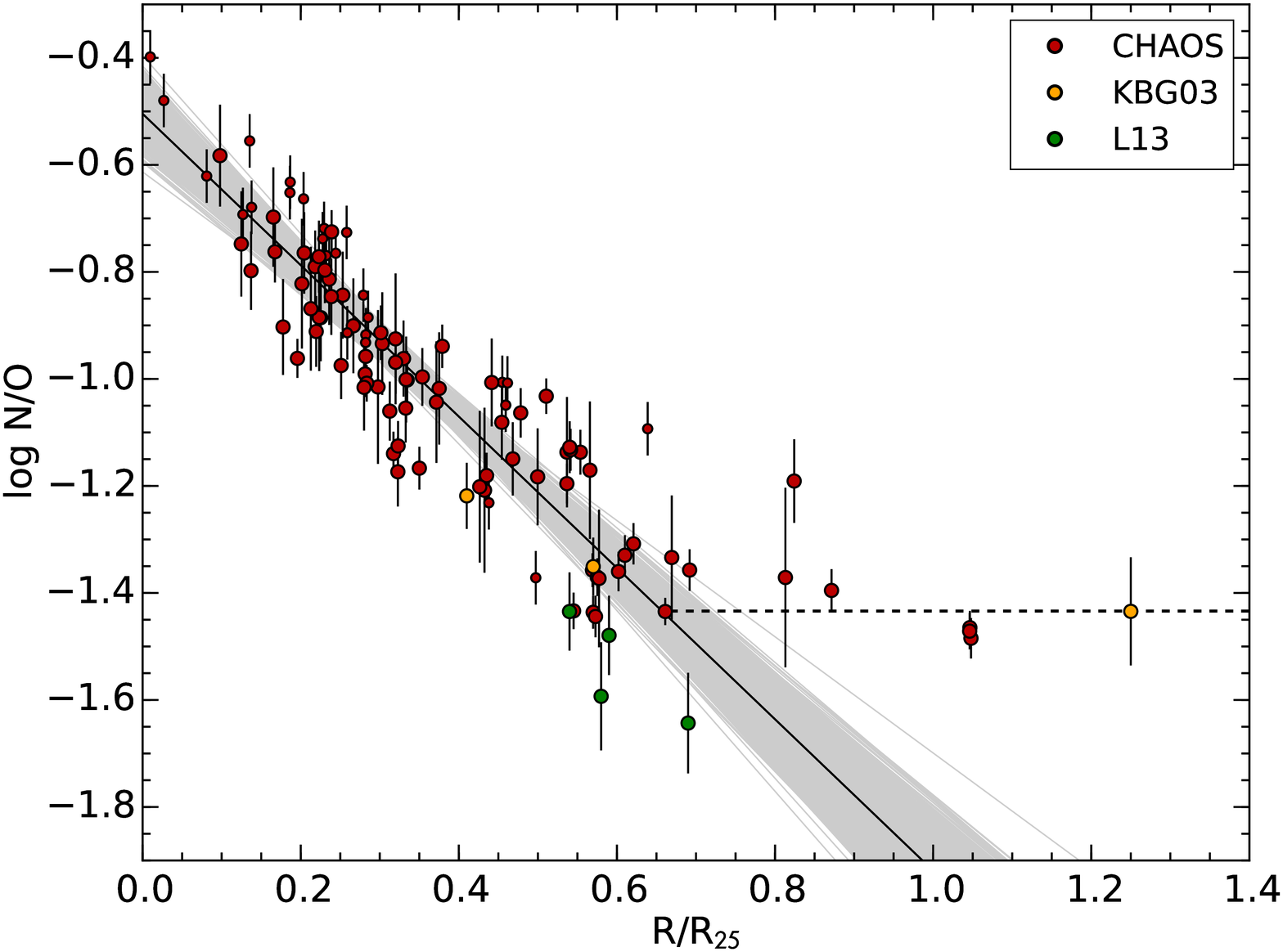}
   \plotone{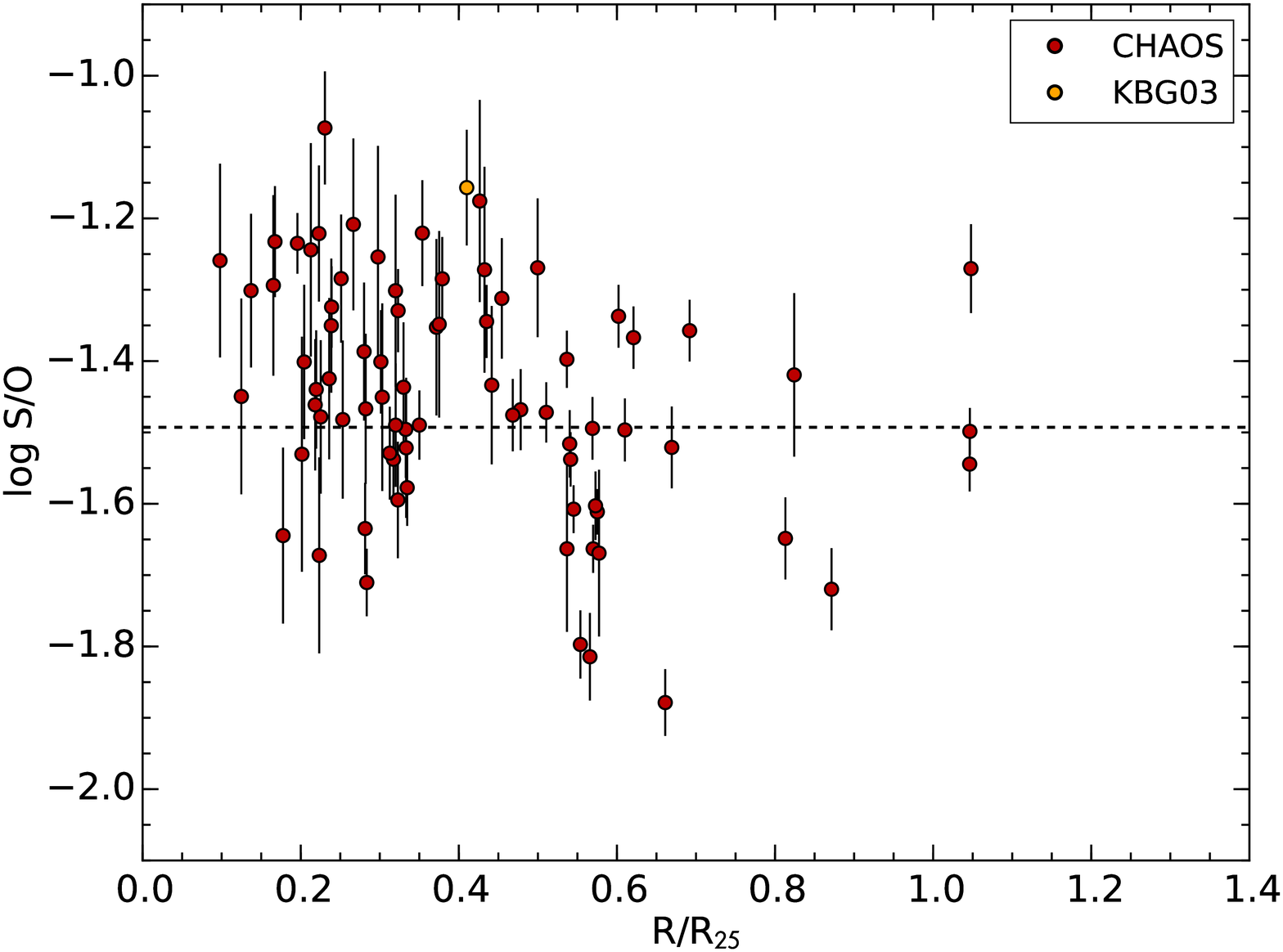}
   \plotone{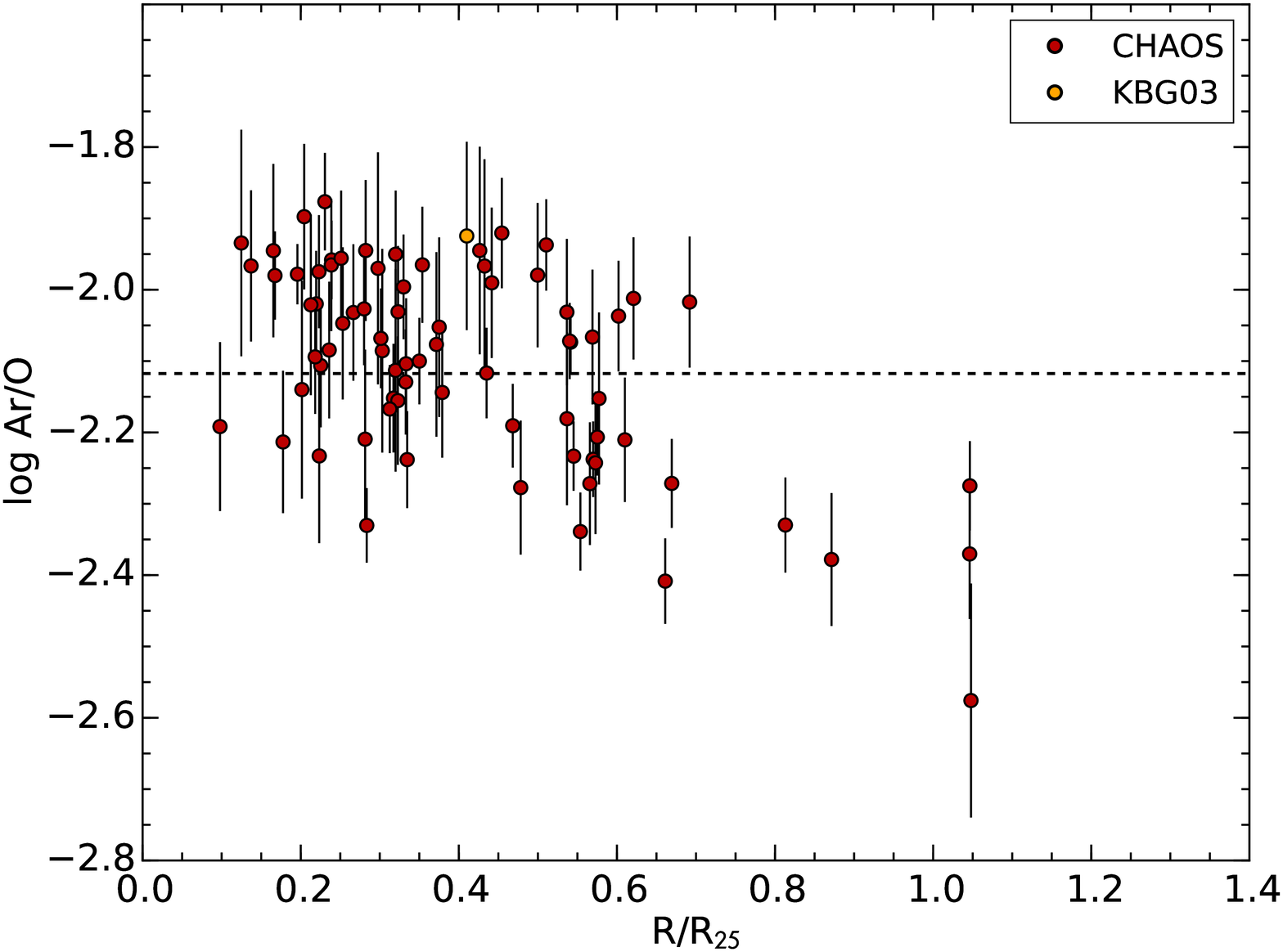}
   \plotone{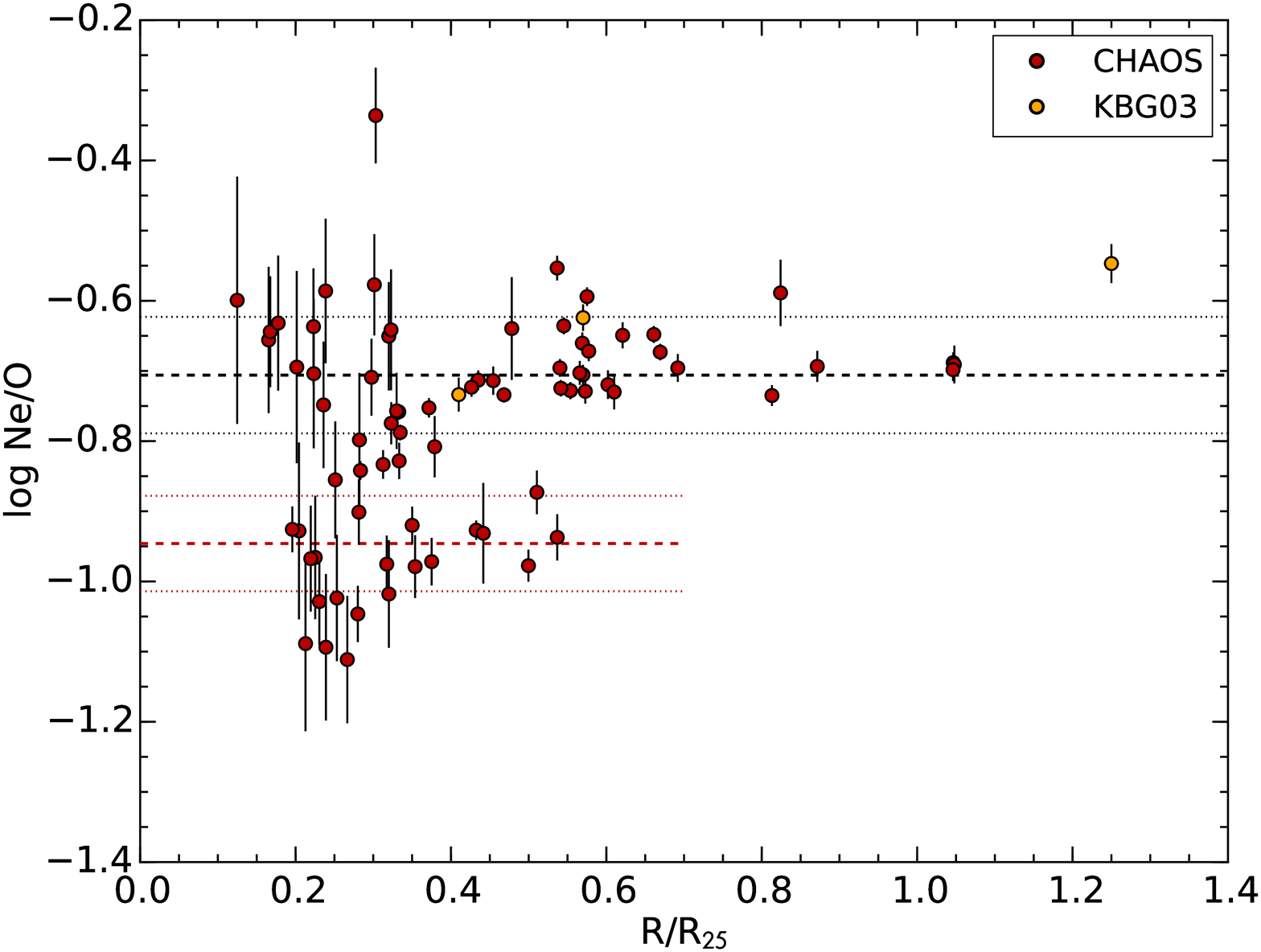}
   \caption{Radial abundance gradients in NGC\,5457.  Points are color-coded by study of origin.  We have recomputed all abundances using identical atomic data and temperature-temperature relations.  We plot the O/H and N/O abundance gradients we derive from direct abundance determinations as solid lines, with the flattened portion of the N/O gradient plotted as a dashed line.  We include data from H\ii\ regions in which auroral lines are not detected as smaller points on the N/O gradient, as this quantity is relatively insensitive to the electron temperature. For the S/O and Ar/O gradients we plot the mean value of the data as a dashed line.  For Ne/O the dashed line indicates the mean of data with a log(Ne/O) greater than $-0.9$.}  
   \label{fig:grad}
\end{figure*}

We calculate the deprojected distance from the galaxy center in units of the isophotal radius, R$_{25}$, using the parameters given in Table \ref{t:n5457global}.  We plot the radial abundance gradients for O/H, N/O, S/O, Ar/O, and Ne/O in Figure \ref{fig:grad}.  The CHAOS observations span the disk of NGC\,5457 from R/R$_{25}\sim$0.10 to R/R$_{25}\sim$1.05, and including previously published direct abundances extends that outward to R/R$_{25}\sim$1.25.  For consistency, we have, in addition to recomputing the abundances using new atomic data as described above, deprojected the distance to each H\ii\ region using the same parameters.  When a given H\ii\ region is present in multiple samples we have elected to adopt the new measurements from this study.  

We note that the N/O abundance ratio is insensitive to changes in temperature because the temperature sensitivities of N$^+$ and O$^+$ are very similar.  Thus, in regions where we do not any detect auroral line, we report the assumed temperature and derived N/O values in Table \ref{t:strongline_no}.  These are based on a semi-empirical approach \citep{vanzee1997} where an electron temperature is used  to calculate the relative N/O abundance ratio. Specifically, we use an electron temperature consistent with the average of the the R-calibration and the S-calibration from \citet{pilyugin2016}.

\begin{deluxetable}{lrr}  
	\tabletypesize{\scriptsize}
	\tablecaption{N/O in NGC\,5457}
	\tablewidth{0pt}
	\tablehead{   
	  \colhead{}	&
	  \colhead{T[N\,\ii]\,(adopted)}	&
	  \colhead{log(N/O)}	\\
	  \colhead{}	&
	  \colhead{(K)}	&
  	  \colhead{(dex)}	
	  }
\startdata
NGC\,5457+134.8+146.0	&		7306	$\pm$	1000	&	-0.72	$\pm$	0.20	\\
NGC\,5457-52.1+41.2	&		6702	$\pm$	1000	&	-0.62	$\pm$	0.20	\\
NGC\,5457-96.2-68.7 	&		7239	$\pm$	1000	&	-0.68	$\pm$	0.20	\\
NGC\,5457-190.6-10.8	&		7642	$\pm$	1000	&	-0.74	$\pm$	0.20	\\
NGC\,5457-139.7-157.6	&		7642	$\pm$	1000	&	-0.77	$\pm$	0.20	\\
NGC\,5457-70.6-92.8   	&		6500	$\pm$	1000	&	-0.56	$\pm$	0.20	\\
NGC\,5457-22.0+1.6	        &		6030	$\pm$	1000	&	-0.48	$\pm$	0.20	\\
NGC\,5457+20.0-104.4	&		7172	$\pm$	1000	&	-0.69	$\pm$	0.20	\\
NGC\,5457-203.8-135.6	&		8180	$\pm$	1000	&	-0.89	$\pm$	0.20	\\
NGC\,5457-205.2-128.7	&		8650	$\pm$	1000	&	-0.92	$\pm$	0.20	\\
NGC\,5457-226.8-77.8	&		8718	$\pm$	1000	&	-0.93	$\pm$	0.20	\\
NGC\,5457+68.2+161.8	&		7038	$\pm$	1000	&	-0.66	$\pm$	0.20	\\
NGC\,5457+44.7+153.7	&		7239	$\pm$	1000	&	-0.65	$\pm$	0.20	\\
NGC\,5457-44.1+149.5	&		6903	$\pm$	1000	&	-0.63	$\pm$	0.20	\\
NGC\,5457-123.1+146.1	&		7508	$\pm$	1000	&	-0.77	$\pm$	0.20	\\
NGC\,5457-192.5+124.7	&		8180	$\pm$	1000	&	-0.84	$\pm$	0.20	\\
NGC\,5457+7.2-3.8	        &		6298	$\pm$	1000	&	-0.40	$\pm$	0.20	\\
NGC\,5457-222.9-366.4	&		9994	$\pm$	1000	&	-1.37	$\pm$	0.20	\\
NGC\,5457-133.9-178.7	&		8314	$\pm$	1000	&	-0.91	$\pm$	0.20	\\
NGC\,5457-66.9-210.8	&		7105	$\pm$	1000	&	-0.73	$\pm$	0.20	\\
NGC\,5457+299.1+464.0	&		9860	$\pm$	1000	&	-1.09	$\pm$	0.20	\\
NGC\,5457-209.1+311.8	&		9591	$\pm$	1000	&	-1.01	$\pm$	0.20	\\
NGC\,5457-219.4+308.7	&		9457	$\pm$	1000	&	-1.05	$\pm$	0.20	\\
NGC\,5457-225.0+306.6	&		8986	$\pm$	1000	&	-1.01	$\pm$	0.20	\\
NGC\,5457-167.8+321.5	&		9860	$\pm$	1000	&	-1.23	$\pm$	0.20
\enddata
  \label{t:strongline_no}
  \tablecomments{N/O abundances for regions where no auroral lines were measured.  This ratio is insensitive to the temperature adopted, nevertheless we report the adopted semi-empirical temperature.}
\end{deluxetable}

Using the Bayesian method described by \citet[][see description in Section 3.1]{kelly2007}, we fit each elemental abundance gradient as both a function of the isophotal radius and the radius in kpc.  We have assumed an error of 0.05\,$\rm{(R/R}_{25}$ on the radial positions, though the effect of including this error was found to be negligible.  This approach allows us a better understanding of the internal dispersion of each element that could be present in the gradient and the error envelope of the slopes.  The best fit to our direct abundance determinations are given as:
\begin{equation} 12 + \rm{log(O/H) = }
	\begin{array}{l}8.716 (\pm0.023) - 0.832 (\pm0.044)~\rm{(R/R}_{25})\\
			       8.715 (\pm0.023) - 0.027 (\pm0.001)~\rm{(R/kpc)}
	\end{array}, 
\end{equation} 
and
\begin{equation} \rm{log(N/O) = }
	\begin{array}{l}-0.505 (\pm0.029) - 1.415 (\pm0.075)~\rm{(R/R}_{25})\\
			       -0.504 (\pm0.028) - 0.046 (\pm0.002)~\rm{(R/kpc)}
	\end{array}.
\end{equation} 

We plot these radial trends as solid lines in Figure \ref{fig:grad}.  We also plot the distribution of possible fits from our Monte-Carlo fitting as lighter gray lines.  For the O/H gradient we also plot the derived gradients of \citet{kennicutt2003} and \citet{li2013} as green and cyan dashed lines respectively.  Both gradients reported in the literature are in agreement within our distribution of uncertainties.  Our fit suggests that the O/H is characterized by an internal dispersion of $0.074\pm0.009$ about the O/H gradient and the N/O gradient is characterized by a dispersion of $0.095\pm0.009$.  While no flattening of the O/H gradient is apparent, past R/R$_{25}$ = 0.64 the N/O is better characterized by a constant value of log(N/O) = $-1.434\pm0.107$.  Finally, as can be seen in Figure \ref{fig:grad} we do not find statistically significant gradients in S/O, Ar/O or Ne/O. 

\subsection{Strong-Line Abundances}
Despite their larger systematic uncertainties \citep[e.g.][]{kewley2008}, strong-line abundance calibrations are often used to determine abundances when auroral lines are not detected.  Even though these strong-line systems lack an absolute calibration \citep[cf.,][]{pilyugin2016} some authors have recently claimed that these methods may be more reliable as they do not rely on a weak line whose measurement uncertainty could be underestimated and thus produce less scatter \citep{ArellanoCordova2016}.

While a full analysis of all strong line methods will be deferred to leverage a complete CHAOS dataset to more fully cover the excitation and abundance parameter space, we have compared the direct abundance results presented in this work with strong-line calibrations.  In particular, we compared with the recent work by \citet{pilyugin2016} that used published observations of NGC\,5457 as the verification set for their strong-line calibration.  Both strong-line calibrations developed by \citet{pilyugin2016} successfully reproduce the abundance gradient within the one-sigma errors.   More importantly, they do not decrease the scatter seen about the gradient.  However, we note that any strong-line calibration can mask actual deviations from the abundance gradient detected by direct abundances.

\section{Discussion}

\subsection{Internal Dispersion and Second Parameters in the Abundance Gradient}
When piecing together the history of chemical enrichment in galaxies, simplifying assumptions are made.  Hence it is common to assume that abundances in a massive spiral galaxy vary radially with little to no scatter azimuthally.  While this assumption must clearly breakdown at some point, it has been hard to quantify the amount of scatter at a given radius given the data that have hitherto been available for most galaxies.   We have measured an internal dispersion of $0.074\pm0.009$ about the O/H gradient in NGC\,5475.  While this is less than the scatter seen by \citet{Rosolowsky2008} (0.11\,dex) and \citet{berg2015} (0.16\,dex), in M33 and NGC\,0628 respectively, it is non-negligible.  
\begin{figure*}[tbp] 
\epsscale{1.2}
   \centering
   \plotone{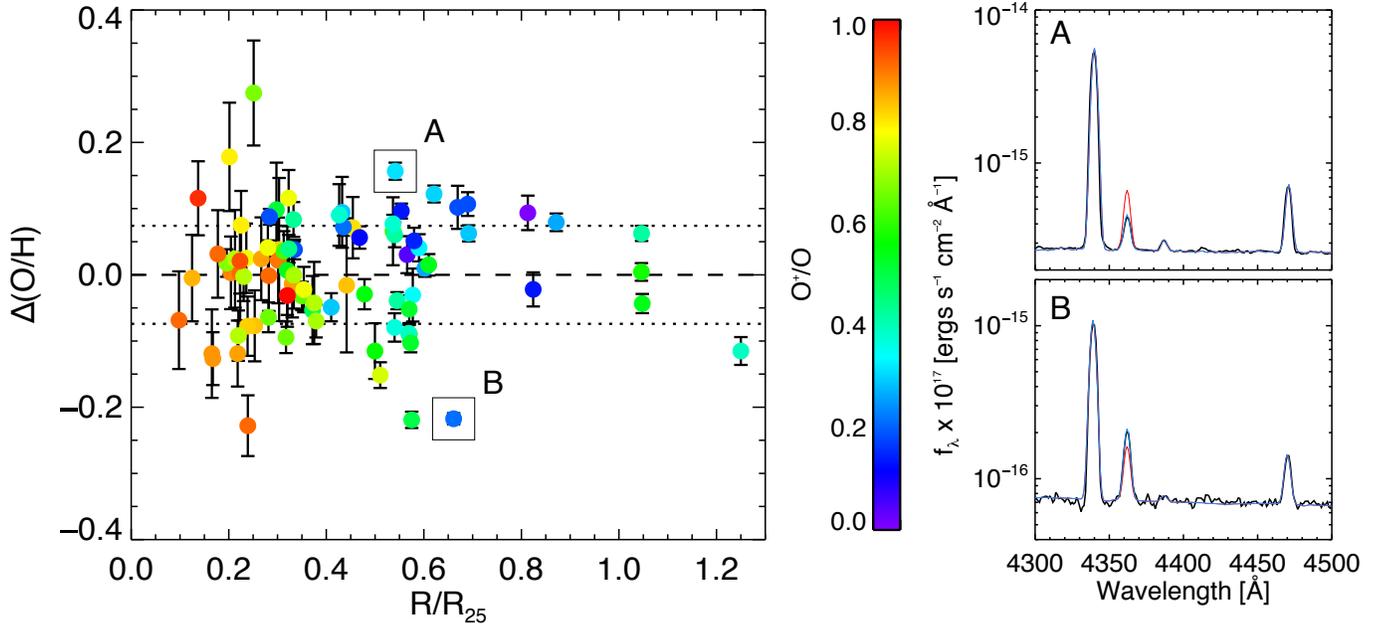}
   \caption{Residuals from the O/H abundance gradient plotted against their R/R25.  Each H\ii\ region is color coded using its derived ratio of O$^+$/O to trace ionization state.   We identify two regions which are located at similar radius and have a similar ionization state but which have very different abundances with residuals on opposite sides of the spread.  Spectral extractions, focused on the portion of the spectrum containing \oiii\,$\lambda$4363, for these two regions are shown to the right with data plotted in black and the modeled spectrum shown in blue.  The redline indicates the measurement of the \oiii\,$\lambda$4363 required to fall on the O/H gradient.}  
   \label{fig:delta}
\end{figure*}  

Although we have gone to great lengths to propagate uncertainties in determining abundances, apparent scatter in the abundances could be due to underestimated uncertainties.  In order to produce an average error equal to the scatter in O/H gradient, we would have to increase the uncertainty of 12+log(O/H) in each H\ii\ region by 0.06\,dex.  Similarly, to reduce the internal dispersion fit by our Monte-Carlo Markov Chain so that it is insignificant we would need to locate an additional 0.1\,dex.  Even though we cannot preclude the possibility of unaccounted for uncertainties (see Section 2.2) we cannot justify an additional 0.06\,dex, much less 0.1\,dex, uncertainty in every abundance determination.

\citet{berg2015} suggest that some of the scatter in their derived O/H gradient could arise from temperature abnormalities.  In contrast, in this dataset we find a good correlation between different ionic temperatures (see Figure \ref{fig:temperature}).  Accordingly, the dispersion seen at a given radius in NGC\,5457 cannot be due to simply using incorrect temperatures.  Furthermore, if scatter were introduced as a function of the electron temperature used to deduce abundances, the dispersion should be suppressed in the N/O ratio, which is significantly less sensitive to the temperature of the nebula.  However, both the O/H and N/O gradients show the approximately the same amount of internal dispersion.  

To better understand the dispersion about the O/H gradient in NGC\,5457 we have investigated it as a function of various additional parameters.  In Figure \ref{fig:delta} we show residuals from the O/H gradient as a function of the radial extent and color-coded based on O$^+$/O.  We also plot dashed lines at $\pm0.074$, the extent of the internal dispersion measured in the O/H.  It is clear that low ionization regions at the center of the galaxy do exhibit both larger uncertainties and an increased dispersion.  However, H\ii\ regions of all excitations fill the extent of the measured internal dispersion.  Additionally, in Figure \ref{fig:delta} we highlight two regions, NGC\,5457-368.3-285.6 and NGC\,5457-540.5-149.9, labeled A and B respectively, which are located at a similar radius in NGC\,5457 and which exhibit similar states of ionization.  However, these two regions lie on opposite sides of the O/H gradient with significant residuals.  To highlight the physical differences in these regions, we also plot a portion of their extracted spectra, focused on the auroral \oiii\,$\lambda$4363.  The blue line shows the modeled spectrum, assuming Gaussian line profiles, while the red line indicates the line strengths needed to yield an abundance that would lie on equation (10).  It is clearly evident that line measurements that would yield abundances consistent with the mean O/H gradient are not consistent with the data. 

In addition to investigating residuals from the O/H gradient with respect to radial location we consider many other logical candidates for possible second parameters.  These include physically motived parameters, such as the ionization parameter P, and data driven considerations, for example the equivalent width of the auroral lines.  In Table \ref{t:correlation} we give the parameters searched and their associated correlation coefficients.  In no cases do any parameters show a significant correlation with the O/H residuals.  
\begin{deluxetable}{lrrr}  
	\tabletypesize{\scriptsize}
	\tablecaption{Possible second parameters in the O/H Abundance Gradient}
	\tablewidth{0pt}
	\tablehead{   
	  \colhead{}	&
  	  \multicolumn{3}{c}{Correlation Coefficients} \\
	  \colhead{}	&
	  \colhead{r}	&
	  \colhead{$\rho$}	&	
	  \colhead{$\tau$}		
	  }
\startdata
P 	 			& 0.22 & 0.29 & 0.19\\
O$^+$/O 			& -0.23 & -0.30 & -0.19 \\
\oiii\,$\lambda$5007 	& 0.21 & 0.27 & 0.18\\
\nii\,$\lambda$6583 	& -0.02 & -0.07 & -0.05\\
H$\alpha$/H$\beta$ 	& 0.13 & 0.12 & 0.08\\
\oiii/\oii\			& 0.25 & 0.29 & 0.19\\
R$_{23}$			& 0.17 & 0.23 & 0.16\\
L$_{H\beta}$		& -0.06 & 0.11 & 0.08\\
EW$_{H\beta}$		& 0.26 & 0.28 & 0.19\\
EW$_{4363}$		& 0.28 & 0.29 & 0.20\\
EW$_{5755}$		& 0.28 & 0.30 & 0.20\\
$\Delta$(N/O)		& -0.12 & 0.07 & 0.11\\
R/R$_{25}$		& -0.01 & 0.07 & 0.04\\
$\phi$			& 0.06 & 0.09 & 0.06
\enddata
  \label{t:correlation}
  \tablecomments{Coefficients from Pearson's (r), SpearmanÕs ($\rho$), and Kendall's ($\tau$) rank correlation of residuals from the O/H gradient and other plausible second parameters. }
\end{deluxetable}

\begin{figure*}[tbp] 
\epsscale{0.579}
   \centering
   \plotone{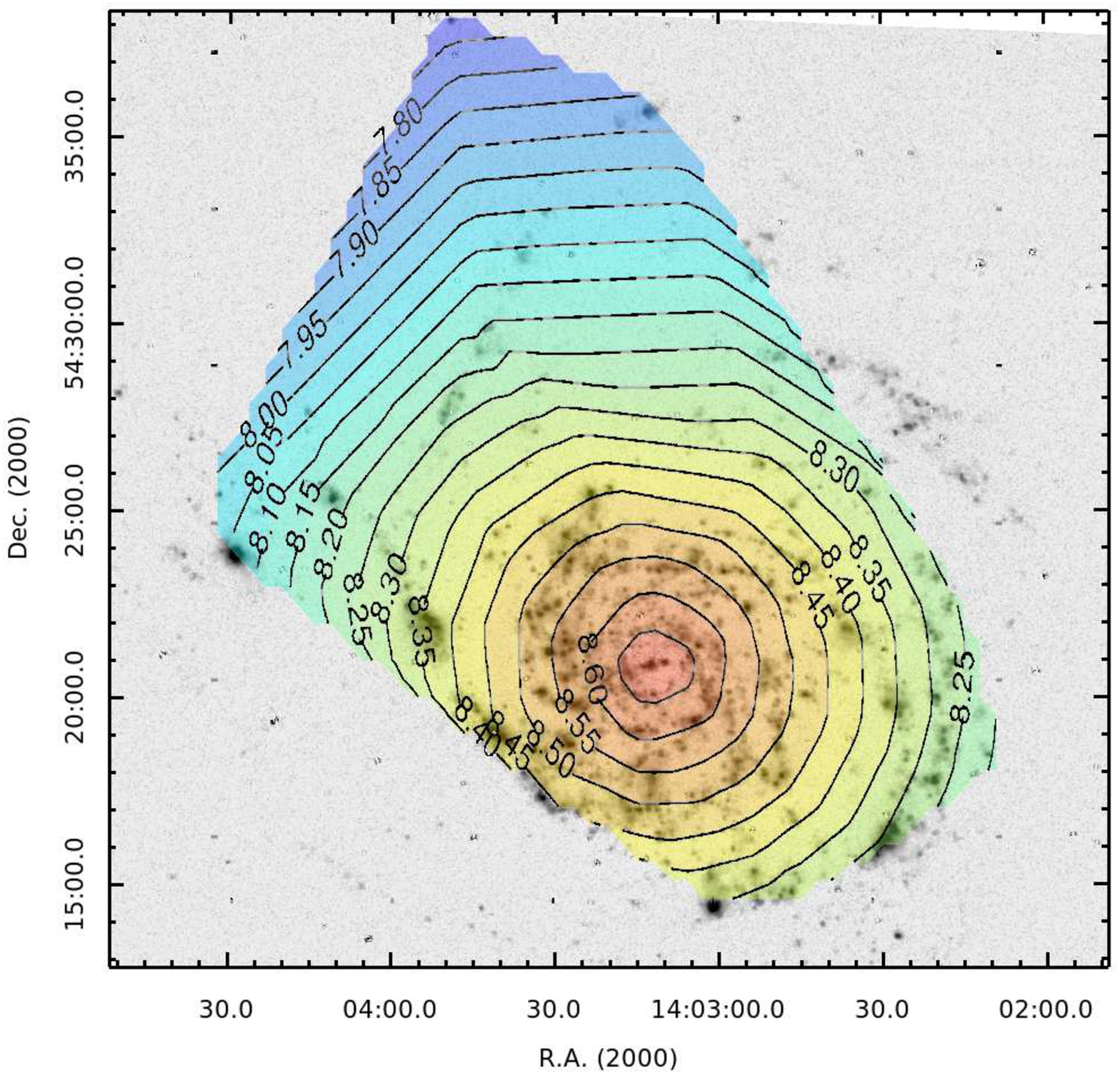}
   \plotone{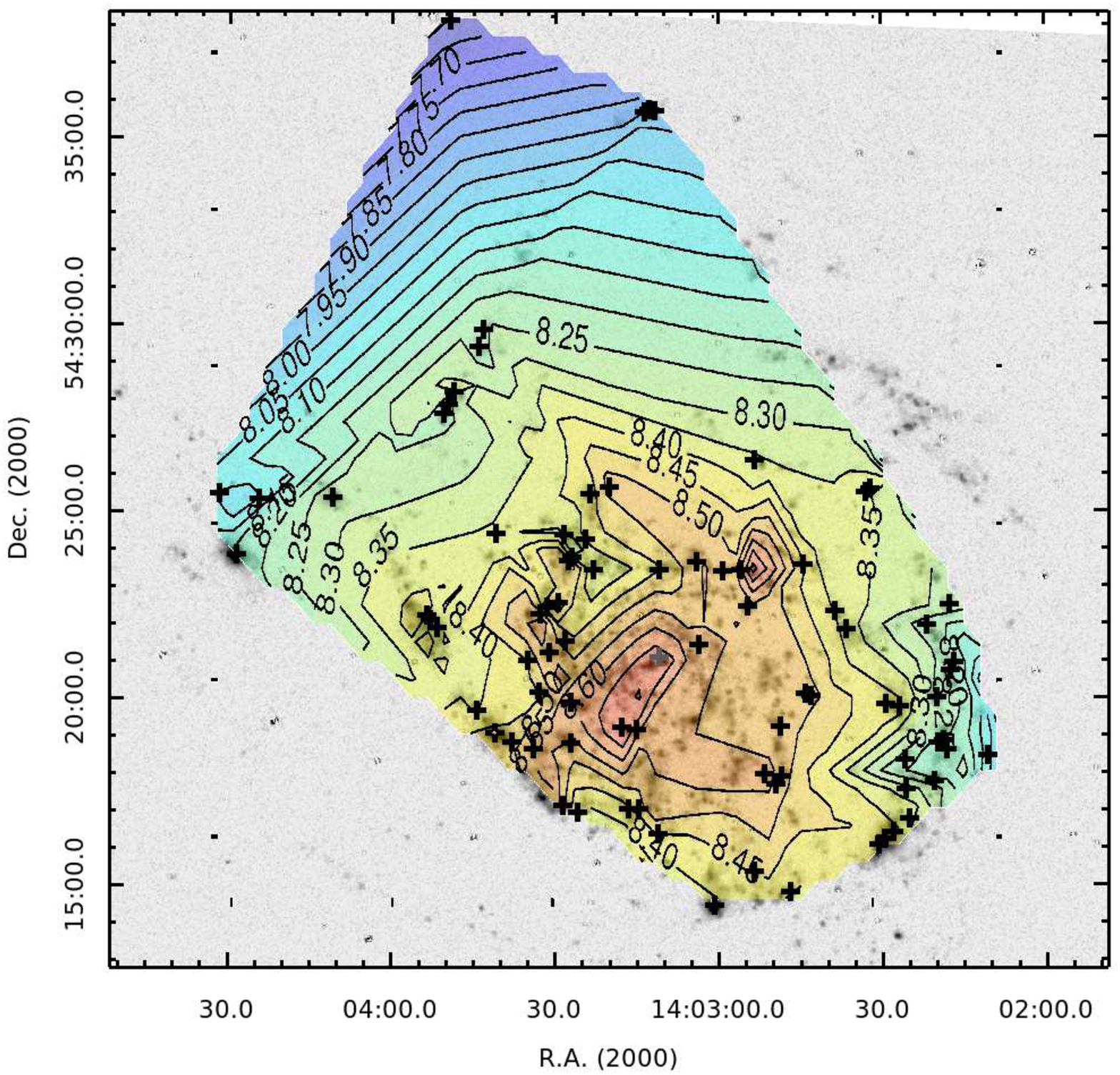}
   \caption{Two dimensional maps of 12+log(O/H) in NGC\,5457.  These maps were created 1) based on assuming a simple projected radial gradient (left) and  2) the ``direct'' abundances measured in regions where auroral lines were detected (right).  Both maps are overlaid on an H$\alpha$ image.  As the number density of direct abundances increase, more structure in gas-phase O/H is clearly apparent.}
   \label{fig:maps}
\end{figure*}

Looking at the two dimensional distribution of abundances in NGC\,5457 we do see a possible mild correlation with the spiral arms, as traced out by the H\ii\ regions.  Using the CONTOUR routine in IDL we interpolate between our irregularly spaced data to create surface maps of oxygen abundance in NGC\,5457.  We show two dimensional maps of 12+log(O/H) in NGC\,5457, based on assuming a simple radial gradient (left) and the direct abundances measured in regions where we detect auroral lines (right), in Figure \ref{fig:maps}.  Comparing these two maps shows that deviations from the O/H gradient are relatively localized.  Indeed, isolating individual spiral arms in fitting abundance gradients does not significantly diminish the internal dispersion measured.  However, care must be taken as comparing the locations of the observed H\ii\ regions (Black $+$ signs) to the structure shows that even with 74 regions, sparse sampling clearly affects the derived map.

While we cannot justify the significant amount of error which must be added to our measurements to eliminate this large amount of scatter, there are other possibilities that remain to be explored.   One likely contribution to this scatter is the fact that we have assumed each H\ii\ region is composed of three zones that exist as isothermal slabs as gas-phase entities embedded in the surrounding interstellar medium.  This is clearly not a physically accurate description of the system.  Rather, as the higher ionization zone is embedded within the lower ionization there must be, at the very least, a gradual transition of electron temperature across this boundary.  Further complicating this picture is the fact that images of H\ii\ regions clearly show they are not smooth Str\"{o}mgren spheres but show noticeable structure.

Effects of this scenario were first discussed by \citet{peimbert1967}, who established a formalism to deal with the inhomogeneity of the temperature zones.  These thermal inhomogeneities have been estimated to necessitate corrections on the order of $0.15-0.35$\,dex \citep{pena2012}, a factor that is larger than the measured dispersion.  While investigation of temperature fluctuations is beyond the scope of this work we note that the MODS spectra contain numerous detections of the Balmer nebular continuum jump, which is useful in determining the normalized standard deviation from the average temperature, i.e, 
\begin{equation}t^2 = \left< \frac{[T(r) - <T(r)>]^2}{<T(r)>^2}\right>.\end{equation}
  Furthermore, the exquisite data will permit detailed modeling by photoionization codes \citep[e.g.,][]{cloudy,mappings} that can go beyond a simple three-zone model.

\subsection{Enrichment relative to Oxygen}
\begin{figure*}[tbp] 
\epsscale{0.8}
   \centering
   \plotone{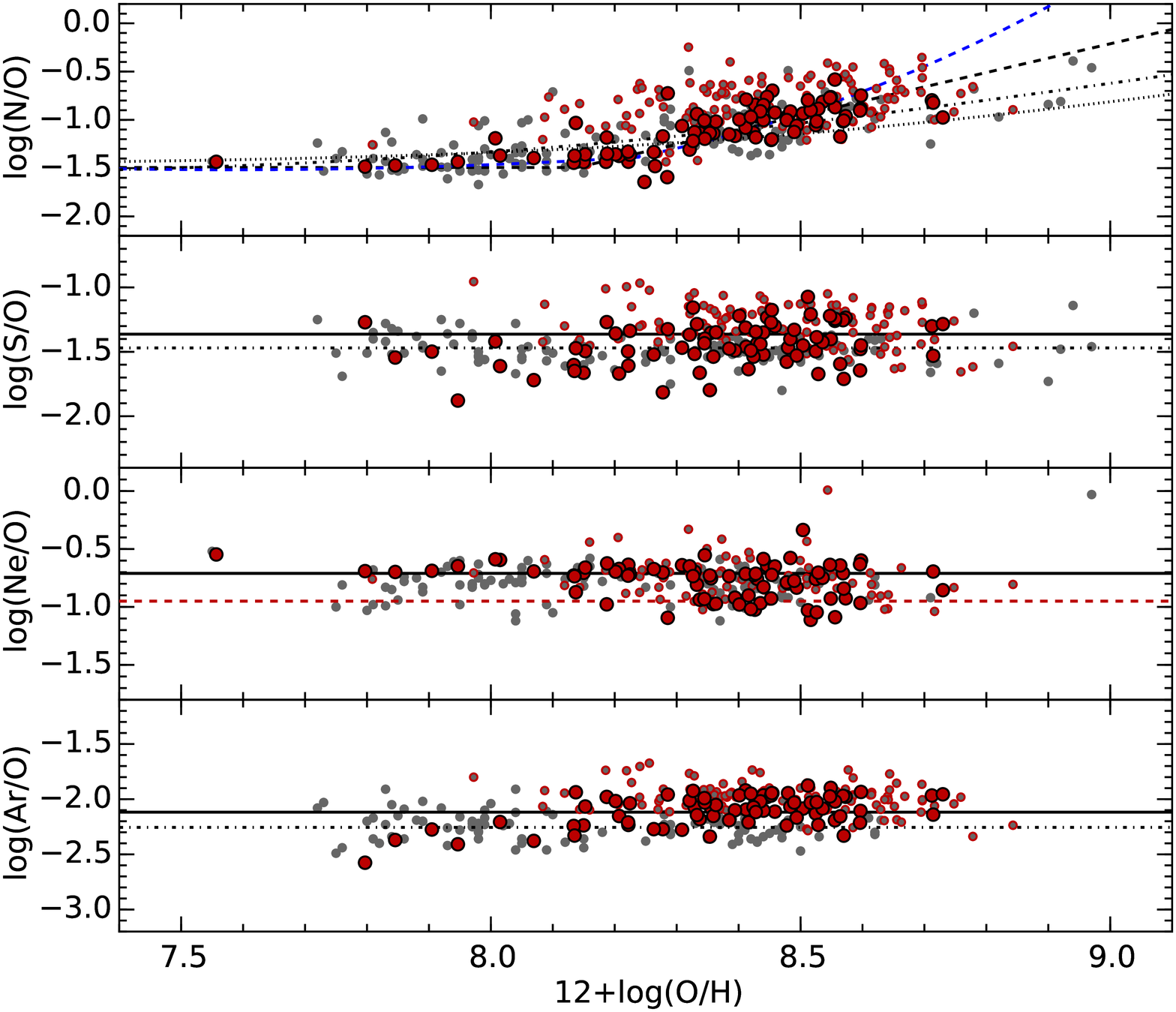}
   \caption{Relative enrichment of nitrogen and the $\alpha$ elements. Light grey circles represent direct abundances taken from \citet{bresolin2005}, \citet{bresolin2009}, \citet{bresolin2009b}, \citep{crox09}, \citet{esteban2009}, \citet{kennicutt2003}, \citet{peimbert2005}, \citet{testor2001}, \citet{testor2003}, \citet{zurita2012}, and \citet{berg2015}.  (Top -- Bottom) log (N/O) as a function of oxygen abundance. The dotted line designates the theoretical curve of \citet{vilacostas}, the black dashed line is the empirical linear fit to galaxies from \citet{pilyugin2010}, the black dot-dash curve is the quadratic fit from \citet{pilyugin2010}, the blue dashed line is the scaled model fit from \cite{groves2004}; log (S/O) as a function of oxygen abundance; log (Ne/O) as a function of oxygen abundance; log (Ar/O) as a function of oxygen abundance.  In the three lower panels, the dashed line in each panel denotes the mean value for our observations of NGC\,5457. Additionally, we have marked the locus of the population of H\ii\ regions with log(Ne/O)$\approx$-1.0 using a dashed red line.}  
   \label{fig:alpha}
\end{figure*}

In Figure \ref{fig:alpha} we plot N/O, S/O, Ne/O, and Ar/O as a function of O/H.  As expected from stellar nucleosynthetic yields \citep[e.g.,][]{woosley1995} and the assumption of a universal initial mass function, these ratios are flat as a function of the radial extent of the galaxy.  The S/O, Ar/O, and Ne/O ratios are largely consistent with the constant values observed in massive and low mass star-forming galaxies.  We note that some of the offset seen between our data (red points) and data culled from the literature is likely due to the adoption of slightly different atomic data and ICFs.  Furthermore we note that by using the updated ICFs we do not detect depressed values of Ar/O and S/O in the most metal rich H\ii\ regions (12 + log(O/H) $\gtrsim$ 8.75) as reported in \citet{croxall2015}.  As seen in other studies, we find increased secondary production of nitrogen at higher O/H  \citep[e.g.,][]{vilacostas}.  Similar to \citet{croxall2015} we detect signatures of massive Wolf-Rayet stars in several apertures (see Table \ref{t:locations}).  However, the presence of features attributed to Wolf-Rayet stars does not appear to lead to any trends of altered abundances.  

While the bulk Ne/O trend is indeed flat, there exists a population of H\ii\ regions with a significant offset to low values of Ne/O, i.e., log(Ne/O)$\approx-1$ (see Figures \ref{fig:grad} and \ref{fig:alpha}).  The existence of two populations of Ne/O, rather than simply a large spread in a single population, is confirmed using a nonparametric kernel technique that does not presuppose an a priori shape to the distribution \citep{vio1994,ryden2006}.  H\ii\ regions that exhibit low Ne/O do not display unusual deficits in other relative abundances, e.g., O/H, N/O, S/O, or Ar/O, and span the full range of O/H we have measured.  These regions are not isolated in a specific location in NGC\,5457, nor are they correlated with the presence of unusual features, such as Wolf-Rayet features or narrow He\,\ii\ emission, in the spectra.  

We note that in our derivation of Ne abundances we have used only the isolated [Ne\,\iii]\,$\lambda$3868 line; thus, the low abundances cannot be attribution to errors in de-blending the [Ne\,\iii]\,$\lambda$3967 line with the nearby H7 line.  Furthermore, Figure \ref{fig:alpha}  shows that similar values of Ne/O are not uncommon in the literature.  However, the reduction in measurement uncertainty permitted by the CHAOS data reveal these regions to be a second population and not simply statistical outliers.  

%%%%%%%%%%%%%%%%%%%%%%%%%%%%%%%%%%%%%%%%%%%%%%%%%%%%%%%%%%%%%%%%
\section {Conclusions}
Leveraging the the power of the LBT with the broad spectral range and sensitivity of MODS, we measure have measured uniform ``direct'' abundances in a large sample of H\ii\ regions in spiral galaxies as part of the CHAOS project.  Here we highlight our observations of 101 H\ii\ regions in NGC\,5457 (M\,101), with a specific focus on the 74 regions where we detected at least one of the temperature sensitive auroral lines of either \oiii, $\lambda4363$, \nii, $\lambda5755$, or \siii, $\lambda6312$.  In 55 cases, more than one of these innately faint emission lines were detected, permitting a comparison of multiple temperature measurements. 

The detection of multiple auroral lines in numerous H\ii\ regions allows us to explore the correlations between the different derived ionic temperature zones.  We have found (1) strong correlations of T[N\,\ii] with T[S\,\iii] and T[O\,\iii], consistent with little or no intrinsic dispersion and the predictions of photoionization modeling; (2) a correlation of T[S\,\iii] with T[O\,\iii], but with significant dispersion and significantly different from the prediction of photoionization modeling; (3) overall agreement between T[N\,\ii], T[S\,\ii], and T[O\,\ii], as expected, but with significant dispersion.

Given these numerous detections of lines, particularly in the less explored low ionization regime, i.e., O$^+$/O$ > 0.4$, we inspect the commonly used ICFs.  Unfortunately, many of the common ICFs are found to be deficient in this region of parameter space.   We propose new empirical ICFs for S and Ar based on our large uniform data set.  

One additional reason we targeted NGC\,5457 was to permit a robust comparison with the data of \citet{kennicutt2003}, \citet{bresolin2007}, and \citet{li2013}.  We show that our pipelines produce results that are in very good agreement with the older data, when updated atomic data are taken into consideration.  Furthermore, our MODS spectra show that some of the unusual temperatures measured in \citet{kennicutt2003} can be rectified with the higher signal-to-noise of our data.  On the other hand, we also detect clear evidence that some unusual temperatures are not due to the difficulty of measuring innately weak lines.  

We measure abundance gradients in both O/H and N/O that are in good agreement with the previous measurements in the literature, namely a steep gradient that drops from approximately solar-neighborhood metallicity in the center of NGC\,5457 \citep{Nieva2012}.  At R/R$_{25}\approx$0.7 we detect a clear flattening of the N/O gradient signaling the low metallicity outer regions of NGC\,5457 are dominated by primary production of N.  We fit these abundance gradients using a Markov-Chain Monte-Carlo approach that includes the possibility that intrinsic dispersion is present.  Despite different dependencies on temperature, both O/H and N/O gradients show the presence of significant internal dispersion at a given radius, namely, $0.074\pm0.009$\,dex and $0.095\pm0.009$\,dex respectively.  Furthermore, we have shown that this dispersion is not related to ionization or other likely second parameters measured from the spectra or their location in the galaxy but appears to be somewhat stochastic. 

In contrast, Ne/O, Ar/O, and S/O all exhibit flat trends with O/H consistent with no evolution of these abundance ratios across the entire disk of NGC\,54557.  While the median value of Ne/O is indeed constant with O/H, we discover a population of H\ii\ regions with a significant offset to low values in Ne/O, $\sim-1$.  This population is present across the range of ionization parameters and abundance trends measured.  Comparing Ne/O measurements from CHAOS with those from the literature show that other galaxies likely have similar populations of low Ne/O, but the number of  observations was too small to robustly detect a separate population.  

As the locations of ongoing star formation, spiral galaxies are where we can observe the chemical evolution of out universe unfolding.  However, only with the advent of very sensitive multi-object spectrographs have we been able to measure the requisite features necessary to truly map out the patterns of the evolution.  These observations are vital in understanding the origin of the elements and their role in the shaping galaxies and the universe as we see it. %\textcolor{red}{So long and thanks for all the fish.}

\acknowledgments
%Funding
K.V.C. is grateful for support from NSF Grant AST-6009233.  K.V.C. is extremely grateful for the chance to have worked with the MODS spectrograph and the CHAOS team.  
%Referee
%MODS
This paper uses data taken with the MODS spectrographs built with funding from NSF grant AST-9987045 and the NSF Telescope System Instrumentation Program (TSIP), with additional funds from the Ohio Board of Regents and the Ohio State University Office of Research.  
%modsIDL
This paper made use of the modsIDL spectral data reduction pipeline developed in part with funds provided by NSF Grant AST-1108693.  
%IDL Builders
We are grateful to D. Fanning, J.\,X. Prochaska, J. Hennawi, C. Markwardt, and M. Williams, and others who have developed the IDL libraries of which we have made use: coyote graphics, XIDL, idlutils, MPFIT, and impro.  
%LBT
This work was based in part on observations made with the Large Binocular Telescope (LBT). The LBT is an international collaboration among institutions in the United States, Italy and Germany. The LBT Corporation partners are: the University of Arizona on behalf of the Arizona university system; the Istituto Nazionale di Astrofisica, Italy; the LBT Beteiligungsgesellschaft, Germany, representing the Max Planck Society, the Astrophysical Institute Potsdam, and Heidelberg University; the Ohio State University; and the Research Corporation, on behalf of the University of Notre Dame, the University of Minnesota, and the University of Virginia.
%NED
This research has made use of the NASA/IPAC Extragalactic Database (NED) which is operated by the Jet Propulsion Laboratory, California Institute of Technology, under contract with the National Aeronautics and Space Administration.

\appendix 
\section{Ionization Correction Factors}

\subsection{Nitrogen}
Nitrogen abundances were derived under the assumption that 
 \begin{equation}
N/O = N^+/O^+.
 \end{equation}
This assumption has been well tested in low abundance nebulae \citep{garnett1990}.  Unfortunately, it's validity is less certain in higher abundance spiral galaxies.  We note that combining observations of the ultraviolet N\,\iii] and the far-IR [N\,\iii] 57$\mu$m line with the now large sample of metal rich H\ii\ regions with optically measured N$^+$/O$^+$ would be greatly beneficial in establishing the veracity of this assumption in cool metal-rich environments.  

We note that the work of \citet{nava2006} suggest that the nitrogen ICF should have a scale factor of 1.08$\pm$0.09 rather than that of unity.  While we have not at this time adopted this scale factor, we note that its adoption does not alter any conclusions and merely results in a simple increase of all reported log(N/O) values by 0.03\,dex.

\subsection{Neon}
Similar to nitrogen, neon can only be measured in one ionization state in optical nebulae.  \citet{peimbert1969} suggested that neon abundance could be derived under the assumption 
 \begin{equation}
Ne/O = Ne^{++}/O^{++}.
 \end{equation}
As neon is significantly stronger in high-ionization nebula, it has predominantly been measured in low abundance nebula.  In pushing towards neon measurements in low-ionization nebula, \citet{kennicutt2003} noted that the scatter in Ne$^{++}$/O$^{++}$ increased at higher O$^{+}$/O.  This could indicate that the neon ionization correction factor does significantly change as H\ii\ regions become cooler.  \citet{dors2013} revisited the ionization fraction of neon by using mid-IR observations of [Ne\,\ii] and [Ne\,\iii] to directly measure the neon ionization fraction.  However, they note that the overlap between visible and IR measurements of neon is limited to more metal poor regions.  

\begin{figure}[tbp] 
\epsscale{0.65}
   \centering
   \plotone{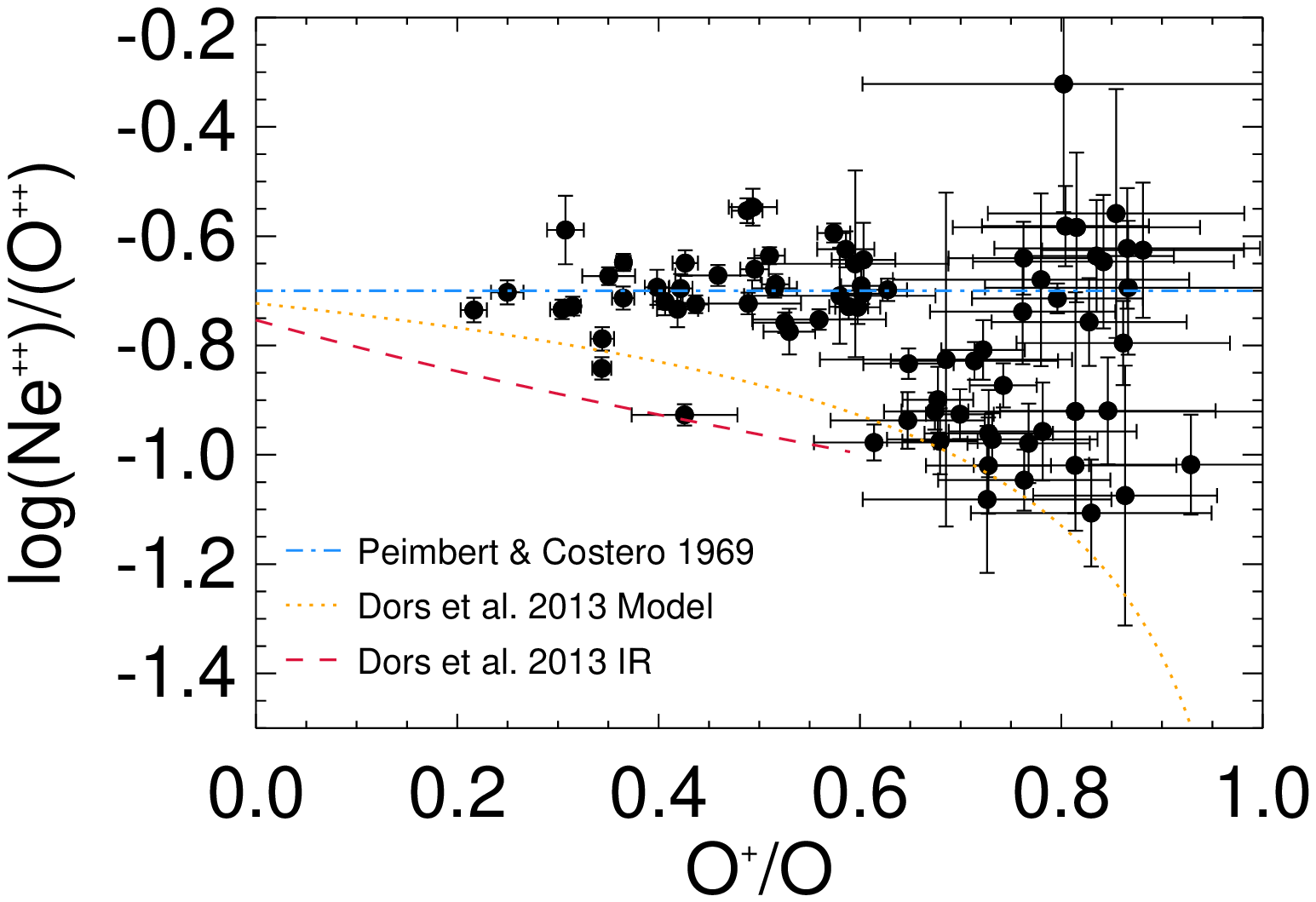}
   \caption{The ratio of Ne$^{++}$/O$^{++}$ as a function of the ionization fraction of oxygen, O$^{+}$/O for H\ii\ regions in NGC\,5457.  We also include the ionization corrections fractions derived by \citet{dors2013}.}  
   \label{fig:Neon_icf}
\end{figure}

Given the sensitivity of the MODS1 spectrograph, we are able to detect the [Ne\,\iii] $\lambda$3869 line in 40 H\ii\ regions with a 12+log(O/H) $\geq$ 8.4.  In Figure \ref{fig:Neon_icf} we show the ratio of Ne$^{++}$/O$^{++}$ as a function of the ionization fraction of oxygen, O$^{+}$/O.  In addition to the \citet{peimbert1969} relation we also plot the two ionization correction factors derived by \citet{dors2013}, 
 \begin{equation}
ICF(Ne^{++})_{model} = 0.741 - 0.08x + 0.393/x,
 \end{equation}
 and
 \begin{equation}
ICF(Ne^{++})_{IR} = 2.382 - 1.301x + 0.05/x (\rm{for~}x > 0.4),
 \end{equation}
 where
  \begin{equation}
  x\equiv O^{++}/(O^+ + O^{++}),
 \end{equation}
 normalized to a constant log(Ne/O) = -0.70.

 The majority of our data are indeed consistent with a constant ratio across a wide range of ionization.  Compared to equation (A2), both ICFs derived by \citet{dors2013} appear to under-predict the amount of Ne relative to oxygen.  Indeed, adopting the ionization correction of  \citet{peimbert1969} for neon yields agreement with the IR measurements of \citet{dors2013}, a constant Ne/O ratio at log(Ne/O) = -0.70.
 
However, we see a related phenomena to the increased scatter at higher O$^{+}$/O reported by \citet{kennicutt2003} in our significantly larger dataset, 42 regions in this portion of parameter space compared to two regions.  Ne$^{++}$/O$^{++}$ appears to bifurcate and inhabit either a high or a low ratio, both of which are flat as a function of O$^{+}$/O.  It is worth noting that the cluster at Ne$^{++}$/O$^{++}\,\sim\,1.0$ appears to agree more closely with the results from IR data \citep{dors2013}.  Given the quality of the spectra, we rule out measurement error as a cause for the low values of Ne$^{++}$/O$^{++}$.  Furthermore, we note that only the [Ne\,\iii] $\lambda$3869 line has been used, thus this can not be attributed to blended lines that have skewed the measurement of the [Ne\,\iii] $\lambda$3967 line, which does give consistent results.

For this work we choose to use the ionization correction formula from \citet{peimbert1969}, i.e., equation (A2).  However, we note that this result is more robust in metal poor H\ii\ regions that are characterized by high ionization parameters.  Indeed, the mid-IR lines in metal rich regions warrant more investigation once the JWST is active.

\subsection{Sulfur}
\begin{figure}[tbp] 
\epsscale{0.55}
   \centering
   \plotone{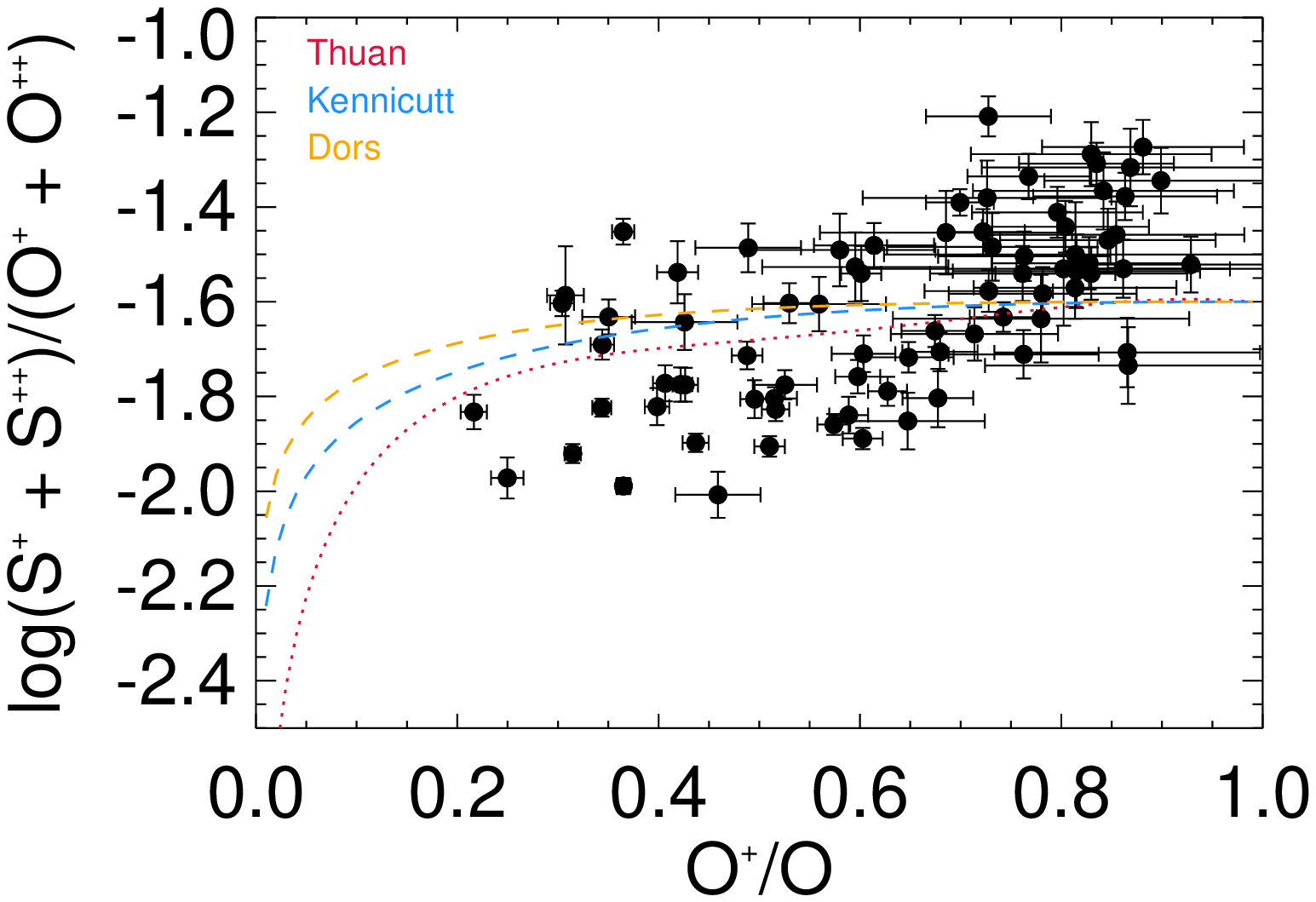}
   \plotone{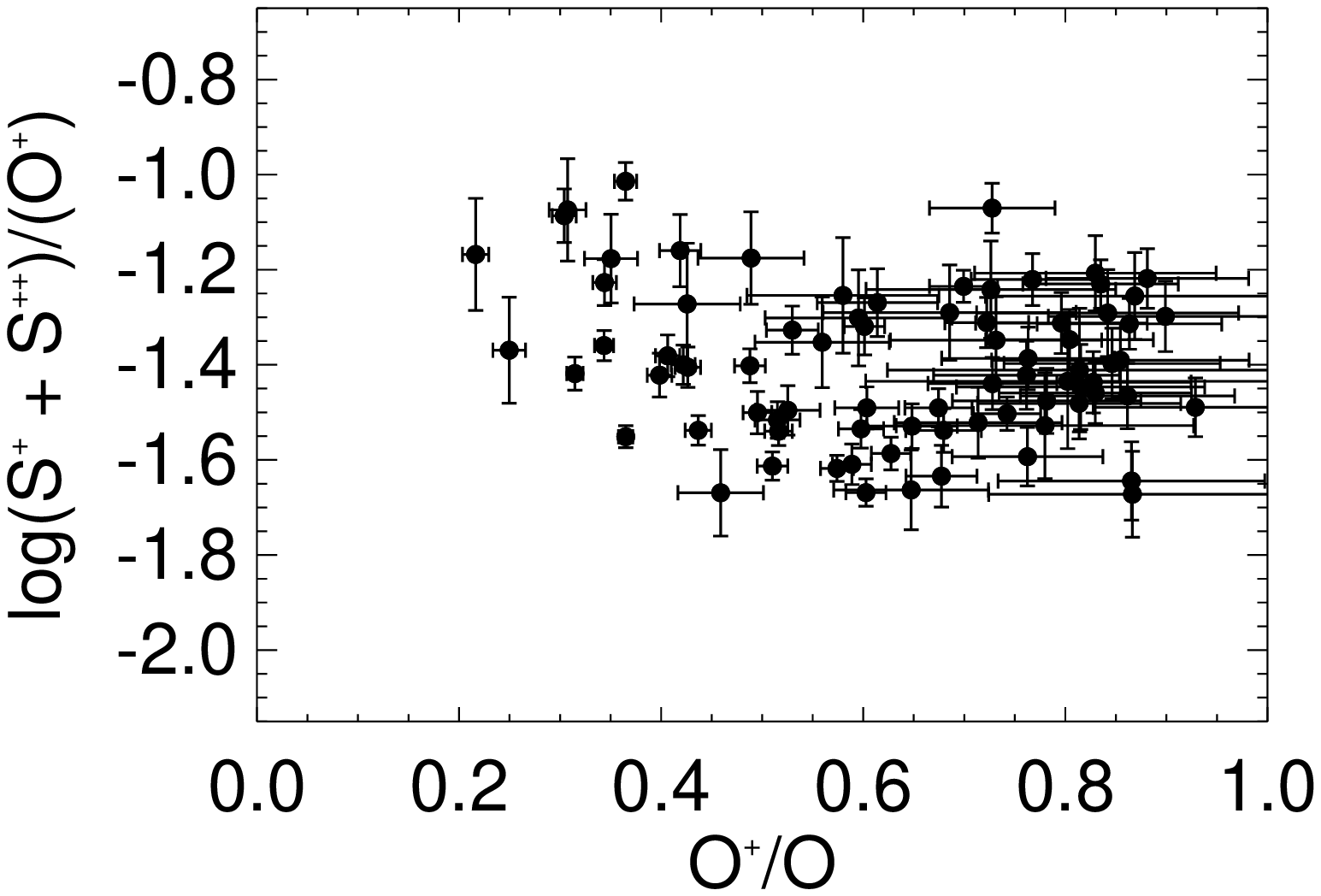}
   \caption{Top: The ratio of (S$^+$ + S$^{++}$)/(O$^+$ + O$^{++}$) as a function of the ionization fraction of oxygen, O$^{+}$/O for H\ii\ regions in NGC\,5457.  We also include the ionization corrections fractions derived by \citet{thuan1995} and \citet{kennicutt2003}.  Bottom: The ratio of (S$^+$ + S$^{++}$)/(O$^+$) as a function of the ionization fraction of oxygen. }  
   \label{fig:sulfur_icf}
\end{figure}

While both S$^+$ and S$^{++}$ are observed in our spectra, we must account for the presence of S$^{+3}$ (30.2\,eV) in the highly ionized region \citep{garnett1989}.  An analytical ionization correction factor for sulfur has been reported by \citet{thuan1995}, based on the photoionization modeling of \citet{stasinska1990},
 \begin{equation}
\frac{S}{S^+ + S^{++}} = \frac{1}{0.013 - 5.10x + x^2[-12.78 + x(14.77-6.11x)]},
 \end{equation}
 where $x$ is again defined in (11).
 This correction is in practice similar to that adopted by  \citet{kennicutt2003} based on earlier photoionization models \citep{stasinska1978}:
 \begin{equation}
\frac{S}{S^+ + S^{++}} = \left[1-\left(1-\frac{O^+}{O}\right)^\alpha \right]^{-1/\alpha},
 \end{equation} 
 with $\alpha=2.5$.  We show these two relations as dashed blue \citep{kennicutt2003} and dotted red \citep{thuan1995} lines, along with our data in Figure \ref{fig:sulfur_icf} for an assumed (constant) intrinsic log(S/O) = $-$1.60.  While our data do not explore the parameter space of highly ionized regions (i.e., O$^+$/O $\leq$0.2), compared to previous works, they more thoroughly populate the low ionization region revealing a clear trend with the oxygen ionization fraction that is not seen in the models.  We also note that atomic data for sulfur has been significantly revised in recent years.  This change in atomic data alone can account for a shift of $\approx$0.1\,dex  between the reported values of (S$^+$ + S$^{++}$)/(O$^+$ + O$^{++}$) and those derived by models using older atomic data.  

Comparing the ionization potentials of the observed sulfur ions, S$^+$ (10.4\,eV) and S$^{++}$ (23.3\,eV), with the observed oxygen ions, O$^+$ (13.6\,eV) and O$^{++}$ (35.1\,eV), suggests that singly ionized oxygen may be a better tracer of the sulfur in cooler H\ii\ regions.  This is clearly illustrated in the bottom panel of Figure \ref{fig:sulfur_icf} where the ratio (S$^+$ + S$^{++}$)/O$^+$ is constant for O$^+$/O $\geq$0.4.  This ratio was previously adopted as an ICF for sulfur \citep{peimbert1969}.  However, \citet{pagel1978} noted a correlation with O$^+$/O in the observed H\ii\ regions at the time.  While we do detect this correlation, the larger dataset, which is in agreement with the data of \citet{pagel1978}, shows that the correlation is limited to high ionization H\ii\ regions.  

Adopting the \citet{thuan1995} or \citet{kennicutt2003} ICFs for sulfur results in a significant trend in S/O with O$^+$/O whereas (S$^+$ + S$^{++}$)/O$^+$ indicates this ratio should be flat \citep{garnett1989}.  Therefore, we adopt  
\begin{equation}
\frac{S}{O} = \frac{S^+ + S^{++}}{O^+},
 \end{equation} 
for O$^+$/O $\geq$0.4 and the ionization correction of \citet{thuan1995} for O$^+$/O $\leq$0.4

\subsection{Argon}
Similar to neon, only one ionization state is observed.  However, like sulfur, in most nebular conditions, thee ionization states can be present: Ar$^+$ in low ionization zones and Ar$^{+3}$ in the highly ionized portions of the nebulae.  Given that the ionization potentials of sulfur and oxygen ions bracket the respective energy levels in argon ions, ratios with both elements have been used to trace the unseen argon.  

\begin{figure}[tbp] 
   \centering
\epsscale{0.55}
   \plotone{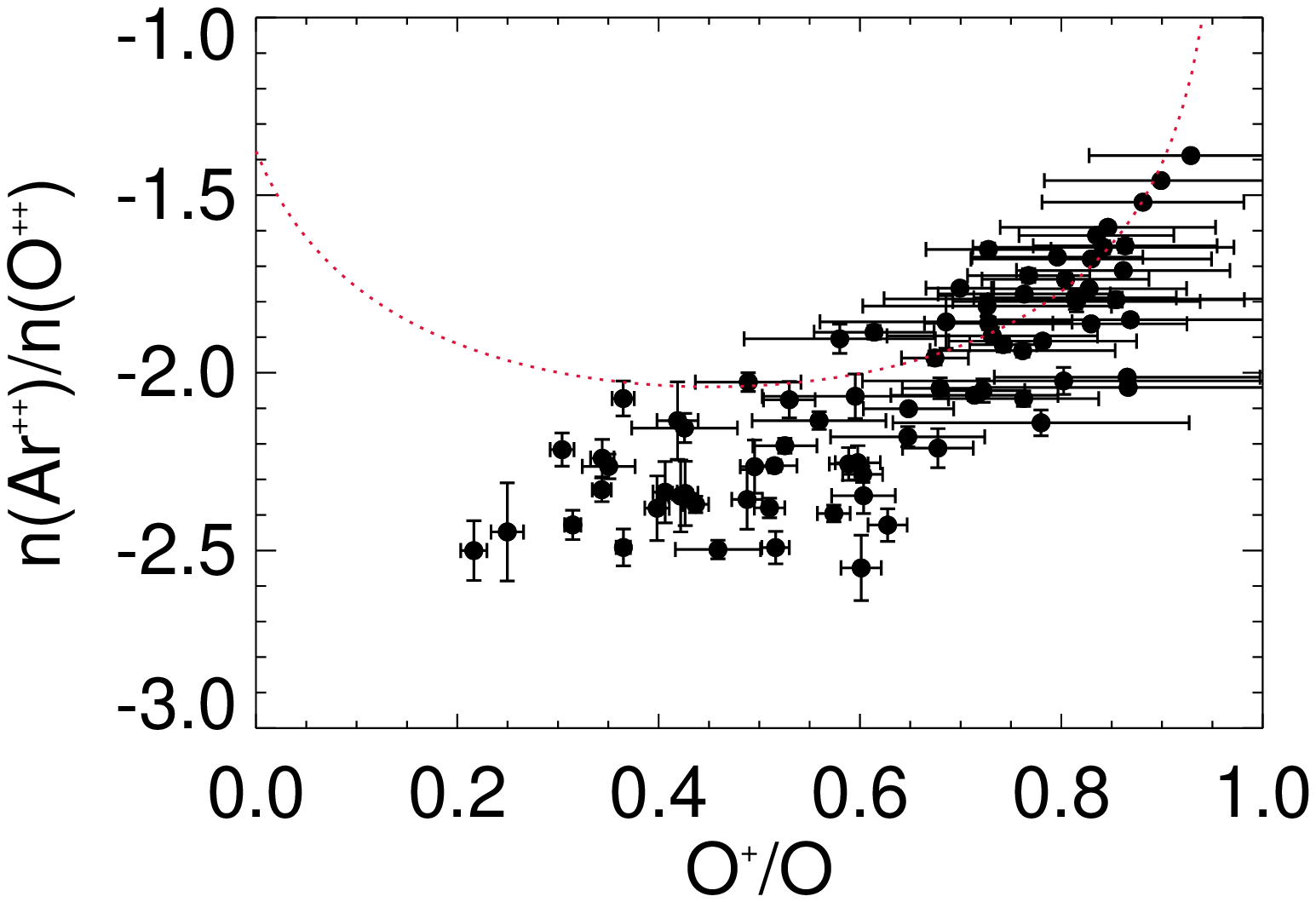}
   \plotone{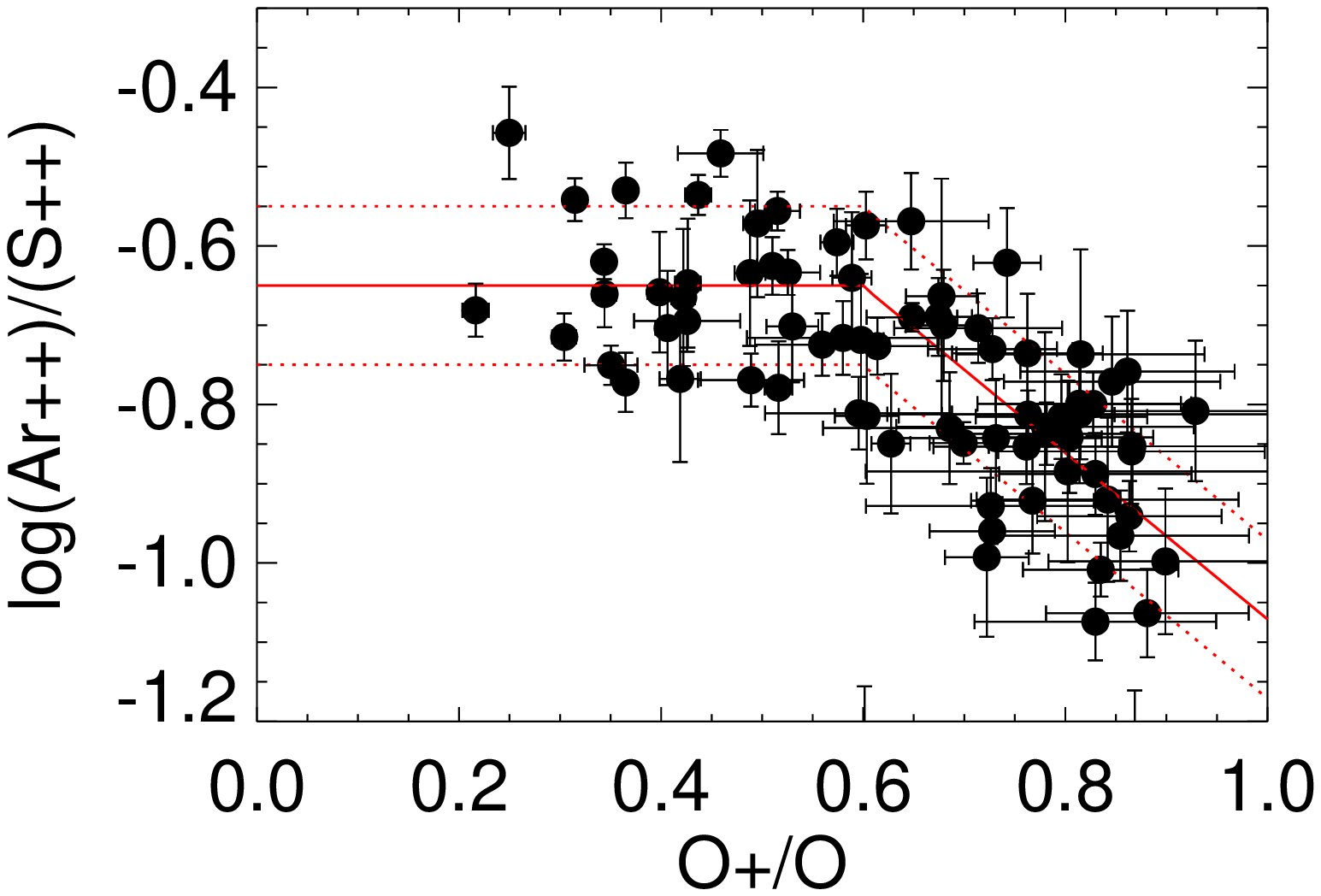}
   \caption{Top: The ratio of Ar$^{++}$/O$^{++}$ as a function of the ionization fraction of oxygen, O$^{+}$/O for H\ii\ regions in NGC\,5457.  We also include the ionization corrections fraction derived by \citet{thuan1995}.  Bottom: The ratio of Ar$^{++}$/O$^++$ as a function of the ionization fraction of oxygen. }  
   \label{fig:argon_icf}
\end{figure}

In Figure \ref{fig:argon_icf} (top panel) we show the ratio of Ar$^{++}$/O$^{++}$ as a function of O$^+$/O along with the ICF of \citet{thuan1995}, 
 \begin{equation}
\frac{Ar}{Ar^{++}} = \frac{1}{0.15 + 2.39x - 2.64x^2}
 \end{equation}
 where $x$ is again defined in (11), for an assumed (constant) intrinsic log(Ar/O) = $-$2.20.  In agreement with \citet{kennicutt2003}, we find that this ratio is strongly correlated with the ionic fraction of oxygen and is thus a poor predictor of Ar/O.  We also show the ratio of Ar$^{++}$/S$^{++}$ as a function of O$^+$/O in Figure \ref{fig:argon_icf} (bottom panel).  \citet{kennicutt2003} found this ratio to be uncorrelated with O$^+$/O and, aside from a single low ionization region, to exhibit a constant value, log(Ar$^{++}$/S$^{++}$)= -0.60.  Our sample clearly shows that low ionization regions do not maintain a constant value of Ar$^{++}$/S$^{++}$ rendering it insufficient as a sole representative of Ar/S.

To correct for the decrease in Ar$^{++}$/S$^{++}$ in low ionization nebula, we adopt a linear correction to Ar$^{++}$/S$^{++}$,
 \begin{equation}
\rm{log}\frac{Ar^{++}}{S^{++}} = -1.049\frac{O^+}{O} -0.022 \mbox{, for } \frac{O^+}{O}  \geq 0.6,
 \end{equation}
to account for an increase in the Ar$^+$ population before adopting this ratio as representative of the Ar/S ratio, with an uncertainty of $\approx\pm0.1$\,dex.  To fully explore the validity of this assumption, observations of the 70\,$\mu$m [Ar\,\ii] line are needed. Currently, this line is only accessible with the SOFIA observatory.

\clearpage 
\LongTables

\begin{deluxetable*}{lccccc|c|c|c|c|c|c|c|c}  
	\tabletypesize{\scriptsize}
	\tablecaption{NGC\,5457 MODS/LBT Observations}
	\tablewidth{0pt}
	\tablehead{   
	  \colhead{H~\ii}	&
	  \colhead{R.A.}		&
	  \colhead{Dec.}	&
	  \colhead{P.A.}	&
	  \colhead{Extraction}	&
	  \colhead{$\frac{R}{R_{25}}$}	&
	  \multicolumn{5}{c}{Auroral Line Detections} &
	  \colhead{}	&
	  \colhead{}	&
	  \colhead{Alt.}		  \\
	  \colhead{Region}	&
	  \colhead{(2000)}	&
	  \colhead{(2000)}	&
	  \colhead{}	&
	  \colhead{(arcsec)}		&
	  \colhead{}	&
	  \colhead{[O\,\iii]}	&
	  \colhead{[N\,\ii]}	&
	  \colhead{[S\,\iii]}	&
	  \colhead{[O\,\ii]}	&
	  \colhead{[S\,\ii]}	&
	  \colhead{WR}	&
	  \colhead{He\,II}	&
	  \colhead{Name}
	  }
\startdata
Total Detections: & & & & & & 50 & 47 & 59 & 67 & 70 & 30 & 10\\ \hline
+7.2-3.8	&	14:03:13.3	&	54:20:52.16	&	85	&	4	&	0.010	&		&		&		&		&		&		&		&	H678	\\
-22.0+1.6	&	14:03:10.0	&	54:20:57.55	&	80	&	8	&	0.027	&		&	\checkmark	&		&		&		&		&		&	H602	\\
-52.1+41.2	&	14:03:06.5	&	54:21:37.23	&	80	&	7	&	0.081	&		&		&		&		&		&		&		&	H555	\\
-75.0+29.3	&	14:03:03.9	&	54:21:25.29	&	80	&	6.5	&	0.098	&		&	\checkmark	&	\checkmark	&		&		&	\checkmark	&		&	H493	\\
+22.1-102.1	&	14:03:15.0	&	54:19:13.93	&	70	&	5	&	0.125	&		&	\checkmark	&		&		&	\checkmark	&		&		&	H699	\\
+20.0-104.4	&	14:03:14.8	&	54:19:11.63	&	70	&	8	&	0.127	&		&		&		&		&	\checkmark	&		&		&	H686	\\
-70.6-92.8	&	14:03:04.4	&	54:19:23.17	&	80	&	8	&	0.135	&		&		&		&		&	\checkmark	&	\checkmark	&		&	H505	\\
+47.9-103.2	&	14:03:18.0	&	54:19:12.82	&	80	&	8	&	0.137	&		&	\checkmark	&		&		&	\checkmark	&		&		&	H760	\\
-96.2-68.7	&	14:03:01.5	&	54:19:47.23	&	80	&	6	&	0.138	&		&		&		&		&		&		&		&	H451	\\
-12.0+139.0	&	14:03:11.1	&	54:23:15.01	&	145	&	7	&	0.165	&		&	\checkmark	&		&	\checkmark	&	\checkmark	&		&		&	H620	\\
+138.9+30.6	&	14:03:28.4	&	54:21:26.49	&	145	&	8	&	0.167	&		&	\checkmark	&	\checkmark	&	\checkmark	&	\checkmark	&	\checkmark	&		&	H972	\\
+134.4-58.8	&	14:03:27.9	&	54:19:57.11	&	70	&	7	&	0.178	&		&	\checkmark	&	\checkmark	&	\checkmark	&	\checkmark	&		&		&	H959	\\
+44.7+153.7	&	14:03:17.6	&	54:23:29.66	&	85	&	4	&	0.186	&		&		&		&		&	\checkmark	&		&		&	H768	\\
-44.1+149.5	&	14:03:07.4	&	54:23:25.50	&	85	&	6	&	0.187	&		&		&		&		&		&		&		&	H567	\\
+164.6+9.9	&	14:03:31.3	&	54:21:05.81	&	145	&	10	&	0.196	&	\checkmark	&	\checkmark	&	\checkmark	&	\checkmark	&	\checkmark	&	\checkmark	& 		&	H1013	\\
+89.3+149.7	&	14:03:22.7	&	54:23:25.66	&	85	&	5	&	0.202	&		&	\checkmark	&	\checkmark	&		&	\checkmark	&		&		&	H864	\\
+68.2+161.8	&	14:03:20.3	&	54:23:37.77	&	85	&	4	&	0.204	&		&		&		&		&	\checkmark	&		&		&	H813	\\
+149.8+92.4	&	14:03:29.6	&	54:22:28.31	&	145	&	3	&	0.204	&		&		&		&	\checkmark	&	\checkmark	&		&		&	H998	\\
-70.2+162.2	&	14:03:04.5	&	54:23:38.19	&	85	&	4	&	0.213	&		&	\checkmark	&	\checkmark	&	\checkmark	&	\checkmark	&		&		&	H504	\\
+166.4+86.3	&	14:03:31.5	&	54:22:22.20	&	145	&	7.5	&	0.218	&		&	\checkmark	&	\checkmark	&	\checkmark	&	\checkmark	&	\checkmark	&		&	H1018	\\
+177.2-42.8	&	14:03:32.8	&	54:20:13.07	&	70	&	6	&	0.220	&		&	\checkmark	&	\checkmark	&	\checkmark	&	\checkmark	&	\checkmark	&		&	H1045	\\
-159.9+89.6	&	14:02:54.2	&	54:22:25.56	&	85	&	8	&	0.223	&		&	\checkmark	&	\checkmark	&	\checkmark	&	\checkmark	&	\checkmark	&		&	H336	\\
+133.1-126.8	&	14:03:27.7	&	54:18:49.17	&	70	&	5	&	0.223	&		&	\checkmark	&	\checkmark	&	\checkmark	&	\checkmark	&	\checkmark	&		&	H949	\\
+177.2+76.1	&	14:03:32.8	&	54:22:12.03	&	145	&	6	&	0.225	&		&	\checkmark	&	\checkmark	&	\checkmark	&	\checkmark	&	\checkmark	&		&	H1040	\\
-190.6-10.8	&	14:02:50.7	&	54:20:45.05	&	80	&	5	&	0.228	&		&		&		&		&		&		&		&	H280	\\
+134.8+146.0	&	14:03:27.9	&	54:23:21.97	&	145	&	2	&	0.230	&		&		&		&		&		&		&		&	H967	\\
-120.2+146.9	&	14:02:58.7	&	54:23:22.84	&	85	&	5	&	0.231	&		&	\checkmark	&	\checkmark	&	\checkmark	&	\checkmark	&	\checkmark	&		&	H399	\\
-123.1+146.1	&	14:02:58.4	&	54:23:22.09	&	85	&	4	&	0.232	&		&		&		&		&		&		&		&	H387	\\
+130.2+157.4	&	14:03:27.4	&	54:23:33.39	&	145	&	7	&	0.236	&		&	\checkmark	&	\checkmark	&	\checkmark	&	\checkmark	&	\checkmark	&		&	H953	\\
+200.3-0.4	&	14:03:35.4	&	54:20:55.42	&	145	&	6.25	&	0.239	&		&		&		&	\checkmark	&		&	\checkmark	&		&	H1071	\\
+129.2+161.7	&	14:03:27.3	&	54:23:37.62	&	145	&	2	&	0.239	&		&		&	\checkmark	&	\checkmark	&	\checkmark	&		&		&	H941	\\
-139.7-157.6	&	14:02:56.5	&	54:18:18.31	&	80	&	10	&	0.244	&		&		&		&		&	\checkmark	&		&	\checkmark	&	H355,NGC5453	\\
-145.1+146.8	&	14:02:55.9	&	54:23:22.72	&	85	&	4	&	0.251	&		&		&	\checkmark	&	\checkmark	&	\checkmark	&		&		&	H351	\\
+103.5+192.6	&	14:03:24.4	&	54:24:08.61	&	145	&	5	&	0.253	&		&	\checkmark	&	\checkmark	&	\checkmark	&	\checkmark	&		&		&	H888	\\
-66.9-210.8	&	14:03:04.9	&	54:17:25.19	&	90	&	3.5	&	0.258	&		&		&		&		&		&		&		&	H510	\\
-133.9-178.7	&	14:02:57.2	&	54:17:57.24	&	90	&	4	&	0.259	&		&		&		&		&		&	\checkmark	&		&	H363,NGC5453	\\
-205.4-98.2	&	14:02:49.0	&	54:19:17.63	&	150	&	4	&	0.267	&		&	\checkmark	&	\checkmark	&	\checkmark	&	\checkmark	&	\checkmark	&		&	H246	\\
-192.5+124.7	&	14:02:50.5	&	54:23:00.57	&	85	&	10	&	0.279	&		&		&		&		&		&		&		&	H284	\\
+17.3-235.4	&	14:03:14.5	&	54:17:00.55	&	90	&	8	&	0.280	&	\checkmark	&	\checkmark	&	\checkmark	&	\checkmark	&	\checkmark	&		&   		&	H694,NGC5458	\\
+36.8-233.4	&	14:03:16.7	&	54:17:02.61	&	70	&	6	&	0.281	&		&	\checkmark	&	\checkmark	&	\checkmark	&	\checkmark	&		&		&	H728,NGC5458	\\
-226.8-77.8	&	14:02:46.6	&	54:19:38.00	&	150	&	12	&	0.282	&		&		&		&		&		&		&		&	H223	\\
+139.0+200.7	&	14:03:28.4	&	54:24:16.62	&	145	&	7	&	0.282	&		&	\checkmark	&		&	\checkmark	&	\checkmark	&		&		&	H969	\\
-205.2-128.7	&	14:02:49.1	&	54:18:47.16	&	150	&	3	&	0.282	&		&		&		&		&		&		&		&	H255	\\
+189.2-136.3	&	14:03:34.1	&	54:18:39.60	&	70	&	4	&	0.284	&	\checkmark	&	\checkmark	&	\checkmark	&	\checkmark	&	\checkmark	&	\checkmark	&		&	H1052	\\
-203.8-135.6	&	14:02:49.2	&	54:18:40.30	&	150	&	4	&	0.285	&		&		&		&		&		&		&		&	H247	\\
-183.9-179.0	&	14:02:51.5	&	54:17:56.90	&	80	&	9	&	0.298	&	\checkmark	&	\checkmark	&	\checkmark	&	\checkmark	&	\checkmark	&		&		&	H290	\\
-249.4-51.3	&	14:02:44.0	&	54:20:04.54	&	80	&	9	&	0.301	&		&		&	\checkmark	&	\checkmark	&	\checkmark	&		&		&	H203	\\
-250.8-52.0	&	14:02:43.8	&	54:20:03.81	&	80	&	3.5	&	0.303	&		&		&	\checkmark	&	\checkmark	&	\checkmark	&		&	\checkmark	&	H203	\\
+225.6-124.1	&	14:03:38.3	&	54:18:51.72	&	70	&	7	&	0.313	&	\checkmark	&	\checkmark	&	\checkmark	&	\checkmark	&	\checkmark	&		&		&	H1086,NGC5461	\\
+117.9-235.0	&	14:03:26.0	&	54:17:00.95	&	70	&	7	&	0.317	&	\checkmark	&		&	\checkmark	&	\checkmark	&	\checkmark	&		&	\checkmark	&	H922	\\
-208.0-180.7	&	14:02:48.7	&	54:17:55.12	&	80	&	10	&	0.320	&	\checkmark	&	\checkmark	&	\checkmark	&	\checkmark	&	\checkmark	&		&		&	H237	\\
-12.3-271.1	&	14:03:11.1	&	54:16:24.95	&	90	&	4	&	0.320	&		&	\checkmark	&	\checkmark	&	\checkmark	&	\checkmark	&	\checkmark	&		&	H618	\\
+142.8-225.2	&	14:03:28.8	&	54:17:10.76	&	70	&	4	&	0.323	&		&		&		&	\checkmark	&	\checkmark	&		&	\checkmark	&	H971	\\
-200.3-193.6	&	14:02:49.6	&	54:17:42.25	&	90	&	16	&	0.323	&	\checkmark	&		&	\checkmark	&	\checkmark	&	\checkmark	&	\checkmark	&		&	H260	\\
+96.7+266.9	&	14:03:23.6	&	54:25:22.90	&	145	&	8	&	0.330	&	\checkmark	&	\checkmark	&	\checkmark	&	\checkmark	&	\checkmark	&	\checkmark	&		&	H875	\\
+67.5+277.0	&	14:03:20.2	&	54:25:33.00	&	145	&	4	&	0.333	&	\checkmark	&	\checkmark	&	\checkmark	&	\checkmark	&	\checkmark	&		&		&	H798	\\
+252.2-109.8	&	14:03:41.3	&	54:19:05.96	&	70	&	4	&	0.333	&	\checkmark	&	\checkmark	&	\checkmark	&	\checkmark	& 	\checkmark	&		&	\checkmark	&	H1100,NGC5461	\\
+254.6-107.2	&	14:03:41.6	&	54:19:08.55	&	70	&	9	&	0.335	&	\checkmark	&	\checkmark	&	\checkmark	&	\checkmark	&	\checkmark &		\checkmark	&		&	H1105,NGC5461	\\
+281.4-71.8	&	14:03:44.7	&	54:19:43.98	&	70	&	8	&	0.350	&	\checkmark	&		&	\checkmark	&	\checkmark	&	\checkmark	&		&	\checkmark	&	H1122	\\
-243.0+159.6	&	14:02:44.7	&	54:23:35.38	&	150	&	4	&	0.354	&	\checkmark	&	\checkmark	&	\checkmark	&	\checkmark	&	\checkmark	&		&		&	H206	\\
+249.3+201.9	&	14:03:41.1	&	54:24:17.71	&	145	&	5	&	0.372	&	\checkmark	&	\checkmark	&	\checkmark	&	\checkmark	&	\checkmark	&		&		&	H1104	\\
-297.7+87.1	&	14:02:38.4	&	54:22:22.84	&	150	&	6	&	0.375	&	\checkmark	&	\checkmark	&	\checkmark	&	\checkmark	&	\checkmark	&		&		&	H185	\\
-309.4+56.9	&	14:02:37.1	&	54:21:52.56	&	150	&	13	&	0.379	&	\checkmark	&		&		&	\checkmark	&		&		&		&	H181,NGC5451	\\
+354.1+71.2	&	14:03:53.0	&	54:22:06.80	&	60	&	7	&	0.426	&	\checkmark	&	\checkmark	&	\checkmark	&	\checkmark	&	\checkmark	&		&		&	H1170,NGC5462	\\
-164.9-333.9	&	14:02:53.7	&	54:15:22.03	&	90	&	6	&	0.433	&	\checkmark	&	\checkmark	&	\checkmark	&	\checkmark	&	\checkmark	&	\checkmark	&		&	H321	\\
+360.9+75.3	&	14:03:53.8	&	54:22:10.81	&	60	&	8	&	0.435	&	\checkmark	&		&	\checkmark	&	\checkmark	&	\checkmark	&		&		&	H1174,NGC5462	\\
-167.8+321.5	&	14:02:53.3	&	54:26:17.43	&	75	&	6	&	0.438	&		&		&		&		&		&		&		&	H295	\\
-178.0+319.1	&	14:02:52.1	&	54:26:14.95	&	75	&	16	&	0.442	&		&		&	\checkmark	&	\checkmark	&	\checkmark	&	\checkmark	&		&	H317	\\
-377.9-64.9	&	14:02:29.3	&	54:19:50.65	&	150	&	13	&	0.454	&	\checkmark	&	\checkmark	&	\checkmark	&	\checkmark	&	\checkmark	&		&		&	H140,NGC5449	\\
-209.1+311.8	&	14:02:48.5	&	54:26:07.69	&	75	&	8	&	0.455	&		&		&		&		&		&	\checkmark	&		&		\\
-219.4+308.7	&	14:02:47.4	&	54:26:04.51	&	75	&	4.5	&	0.459	&		&		&		&		&		&		&		&		\\
-225.0+306.6	&	14:02:46.7	&	54:26:02.45	&	75	&	4	&	0.461	&		&		&		&		&		&		&		&	H234	\\
-99.6-388.0	&	14:03:01.1	&	54:14:27.97	&	90	&	5	&	0.468	&	\checkmark	&	\checkmark	&	\checkmark	&	\checkmark	&	\checkmark	&	 \checkmark	&		&	H416,NGC5455	\\
-397.4-71.7	&	14:02:27.1	&	54:19:43.76	&	124	&	11	&	0.478	&	\checkmark	&	\checkmark	&		&		&		&		&		&	H115,NGC5449	\\
-222.9-366.4	&	14:02:47.1	&	54:14:49.40	&	90	&	2	&	0.497	&		&		&		&		&	\checkmark	&		&		&	H231	\\
-226.9-366.4	&	14:02:46.6	&	54:14:49.40	&	90	&	7	&	0.500	&	\checkmark	&	\checkmark	&	\checkmark	&	\checkmark	&	\checkmark	&		&		&	H219	\\
-405.5-157.7	&	14:02:26.2	&	54:18:17.79	&	124	&	10	&	0.511	&	\checkmark	&		&		&	\checkmark	&	\checkmark	&	\checkmark	&		&	H104	\\
-345.5+273.8	&	14:02:32.9	&	54:25:29.41	&	75	&	3	&	0.537	&	\checkmark	&		&	\checkmark	&	\checkmark	&	\checkmark	&		&		&	H167	\\
-410.3-206.3	&	14:02:25.6	&	54:17:29.09	&	124	&	4	&	0.537	&	\checkmark	&	\checkmark	&	\checkmark	&	\checkmark	&	\checkmark	&	\checkmark	&		&	H103	\\
-371.1-280.0	&	14:02:30.1	&	54:16:15.53	&	124	&	8	&	0.540	&	\checkmark	&	\checkmark	&	\checkmark	&	\checkmark	&	\checkmark	&	 \checkmark	&		&	H143,NGC5447	\\
-368.3-285.6	&	14:02:30.5	&	54:16:09.89	&	124	&	9	&	0.541	&	\checkmark	&	\checkmark	&	\checkmark	&	\checkmark	&	\checkmark	&	 \checkmark	&		&	H149,NGC5447	\\
-455.7-55.8	&	14:02:20.4	&	54:19:59.55	&	124	&	17	&	0.545	&	\checkmark	&		&	\checkmark	&	\checkmark	&	\checkmark	&		&		&	H59	\\
-392.0-270.1	&	14:02:27.8	&	54:16:25.34	&	124	&	13	&	0.554	&	\checkmark	&	\checkmark	&	\checkmark	&	\checkmark	&	\checkmark	&	 \checkmark	&		&	H128,NGC5447	\\
-414.1-253.6	&	14:02:25.2	&	54:16:41.84	&	124	&	4	&	0.566	&	\checkmark	&	\checkmark	&	\checkmark	&	\checkmark	&	\checkmark	&		&		&	H98	\\
-464.7-131.0	&	14:02:19.4	&	54:18:44.25	&	124	&	4	&	0.569	&	\checkmark	&		&	\checkmark	&	\checkmark	&	\checkmark	&		&		&	H51	\\
-466.1-128.2	&	14:02:19.2	&	54:18:47.06	&	124	&	4	&	0.570	&	\checkmark	&		&	\checkmark	&	\checkmark	&	\checkmark	&		&		&	H46	\\
-479.7-3.9	&	14:02:17.6	&	54:20:51.33	&	124	&	4	&	0.573	&	\checkmark	&		&	\checkmark	&	\checkmark	&		&		&		&	H35	\\
-481.4-0.5	&	14:02:17.4	&	54:20:54.75	&	124	&	4	&	0.575	&	\checkmark	&		&	\checkmark	&	\checkmark	&	\checkmark	&	\checkmark	&		&	H27	\\
-453.8-191.8	&	14:02:20.7	&	54:17:43.47	&	124	&	14	&	0.577	&	\checkmark	&	\checkmark	&	\checkmark	&	\checkmark	&	\checkmark	&		&		&	H71	\\
+331.9+401.0	&	14:03:50.6	&	54:27:36.60	&	153.7	&	8	&	0.602	&	\checkmark	&		&		&	\checkmark	&		&		&		&	H1151	\\
+324.5+415.8	&	14:03:49.7	&	54:27:51.46	&	153.7	&	9	&	0.610	&	\checkmark	&		&		&		&		&		&		&	H1148	\\
+315.3+434.4	&	14:03:48.7	&	54:28:10.10	&	153.7	&	4	&	0.621	&	\checkmark	&		&		&		&		&		&		&	H1146	\\
+299.1+464.0	&	14:03:46.8	&	54:28:39.70	&	153.7	&	5	&	0.639	&		&		&		&		&		&		&		&	H1137	\\
-540.5-149.9	&	14:02:10.7	&	54:18:25.13	&	124	&	9	&	0.661	&	\checkmark	&		&	\checkmark	&	\checkmark	&	\checkmark	&		&		&	H8	\\
+509.5+264.1	&	14:04:10.9	&	54:25:19.20	&	87.5	&	16	&	0.669	&	\checkmark	&	\checkmark	&	\checkmark	&	\checkmark	&	\checkmark	&		&	\checkmark	&	H1216	\\
+266.0+534.1	&	14:03:43.0	&	54:29:49.89	&	153.7	&	8	&	0.692	&	\checkmark	&		&		&	\checkmark	&		&	\checkmark	&		&	H1125	\\
+667.9+174.1	&	14:04:29.0	&	54:23:48.56	&	90	&	4	&	0.813	&	\checkmark	&	\checkmark	&	\checkmark	&	\checkmark	&	\checkmark & 			& 	\checkmark 	&	H1239,NGC5471	\\
+650.1+270.7	&	14:04:27.0	&	54:25:25.29	&	87.5	&	5	&	0.824	&	\checkmark	&		&		&		&		&		&	\checkmark	&	H1231	\\
+692.1+272.9	&	14:04:31.8	&	54:25:27.24	&	87.5	&	4	&	0.871	&	\checkmark	&		&	\checkmark	&	\checkmark	&		&		&		&	H1248	\\
+1.0+885.8	&	14:03:12.6	&	54:35:41.83	&	90	&	4	&	1.046	&	\checkmark	&		&		&		&		&		&	\checkmark	&	H681	\\
+6.6+886.3	&	14:03:13.3	&	54:35:42.29	&	90	&	21	&	1.046	&	\checkmark	&		&		&	\checkmark	&		&		&		&	H641	\\
-8.5+886.7	&	14:03:11.5	&	54:35:42.70	&	90	&	8	&	1.048	&	\checkmark	&		&		&		&		&		&		&	H672			
\enddata
	\label{t:locations}
	\tablecomments{Observing logs for H\ii\ regions observed in NGC 5457 using MODS on the LBT during the 2015A semester.   Each exposure was taken with an integrated exposure time of 1200s on clear nights, with, on average $\sim$1\farcs00 seeing, and airmasses less than 1.3.  Slit ID, composed of the offset in R.A. and Dec., in arcseconds, from the central position listed in Table \ref{t:n5457global} is listed in Column 1. The right ascension and declination of the individual H\ii\ regions are given in units of hours, minutes, and seconds, and degrees, arcminutes, and arcseconds respectively in columns 2 and 3.   We list the position angle of the slit and the size of the region extracted along each slit in columns 4 and 5.  The de-projected distances of H\ii\ regions from the center of the galaxy as a fraction of R$_{25}$ is listed in column 6.  Columns 7-11 highlight which regions have [O\,\iii] $\lambda$4363, [N\,\ii] $\lambda$5755, [S\,\iii] $\lambda$6312, [O\,\ii] 7330, [S\,\ii] $\lambda$4070 auroral lines detections at the 3$\sigma$ significance level.  We note in columns 12 and 13 which H\ii\ regions have detections of broad Wolf-Rayet features and narrow He\,\ii\,$\lambda$4686 emission.  Finally Column 14 reports alternate names for the observed H\ii\ regions primarily taken from \citet{hodge1990}}
\end{deluxetable*}

\clearpage
\begin{deluxetable*}{lcccccccccc}  
	\tabletypesize{\scriptsize}
	\tablecaption{Line fluxes relative to H$\beta$}
	\tablewidth{0pt}
	\tablehead{   
	  \colhead{Ion}	&
	  \colhead{NGC5457+7.2-3.8}		&
	  \colhead{NGC5457-22.0+1.6}		&
	  \colhead{NGC5457-52.1+41.2}	&
	  \colhead{NGC5457-75.0+29.3}	&
	  \colhead{NGC5457+22.1-102.1}		
}
\startdata
     \multicolumn{6}{c}{\includegraphics[height=60mm]{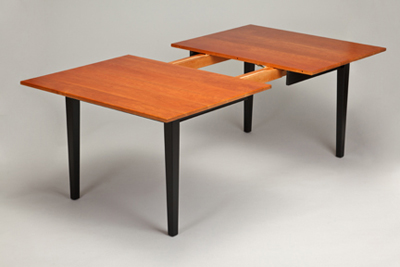}}
\enddata
	\label{t:lineflux}
	\tablecomments{Table will be available in the accepted paper.\\}
\end{deluxetable*}

\clearpage

\begin{deluxetable*}{cccccccc}  
	\tabletypesize{\scriptsize}
	\tablecaption{Abundances in NGC\,5457}
	\tablewidth{0pt}
	\tablehead{   
	  \colhead{}	&
	  \colhead{NGC5457-75.0+29.3 }	&
	  \colhead{NGC5457+22.1-102.1 }	&
	  \colhead{NGC5457+47.9-103.2}	&
	  \colhead{NGC5457-12.0+139.0 }	&
	  \colhead{NGC5457+138.9+30.6}	
	  }
\startdata
     \multicolumn{6}{c}{\includegraphics[height=60mm]{table.png}}
\label{t:labundances}
	\tablecomments{Table will be available in the accepted paper.\\}
\end{deluxetable*}

\end{document}